\documentclass{aa}  

\usepackage{graphicx}
\usepackage{txfonts}
\usepackage{color}

\def\mstar  {$M_{\star}$}
\def\macc   {$\dot{M}_{\rm acc}$}

\def\lacc   {$L_{\rm acc}$}

\def\msun {$M_{\odot}$}

\def\lstar {$L_\star$}

\def\lline {$L_{\rm line}$}

\def\laccnoise   {$L_{\rm acc,noise}$}

\def\nodata {...}
\newcommand{\teff}{$T_{\rm eff}$}
\newcommand{\logg}{$\log g$}
\newcommand{\vsini}{$v\sin i$}

\defcitealias{HH14}{HH14}
\defcitealias{Manara13a}{MTR13}

\begin{document}

   \title{An extensive VLT/X-Shooter library of photospheric templates \\of pre-main sequence stars\thanks{Based on observations made with ESO Telescopes at the La Silla Paranal Observatory under programme ID 096.C-0979}}

\titlerunning{Photospheric templates of PMS stars}
\authorrunning{C.F. Manara, A. Frasca, J.M. Alcal\'a et al.}

    \author{C.F. Manara \inst{1}\fnmsep\thanks{ESA Research Fellow}, A.~Frasca\inst{2}, J.M.~Alcal\'a\inst{3}, A.~Natta\inst{4,5}, B.~Stelzer\inst{6,7}, L.~Testi\inst{8,5,9}
          }

   \institute{Scientific Support Office, Directorate of Science, European Space Research and Technology Centre (ESA/ESTEC), Keplerlaan 1, 2201 AZ Noordwijk, The Netherlands \\
              \email{cmanara@cosmos.esa.int}
\and
INAF - Osservatorio Astrofisico di Catania, via S. Sofia, 78, 95123 Catania, Italy
\and 
INAF-Osservatorio Astronomico di Capodimonte, via Moiariello 16, 80131 Napoli, Italy
\and
School of Cosmic Physics, Dublin Institute for Advanced Studies, 31 Fitzwilliams Place, Dublin 2, Ireland
         \and
         INAF-Osservatorio Astrofisico di Arcetri, L.go E. Fermi 5, 50125 Firenze, Italy
\and
Eberhard Karls Universit\"at, Kepler Center for Astro and Particle Physics, Institut f\"ur Astronomie und Astrophysik, Sand 1, 72076 T\"ubingen, Germany
\and 
INAF - Osservatorio Astronomico di Palermo, Piazza del Parlamento 1, 90134 Palermo, Italy
\and
             European Southern Observatory, Karl-Schwarzschild-Strasse 2, 85748 Garching bei M\"unchen, Germany
         \and
             Excellence Cluster Universe, Boltzmannstr. 2, 85748 Garching, Germany
             }

   \date{Received March 17, 2017; accepted May 28, 2017}

% \abstract{}{}{}{}{} 
% 5 {} token are mandatory
 
  \abstract
  % context heading (optional)
  % {} leave it empty if necessary  
   {Studies of the formation and evolution of young stars and their disks rely on knowledge of the stellar parameters of the young stars. The derivation of these parameters is commonly based on comparison with photospheric template spectra. Furthermore, chromospheric emission in young active stars impacts the measurement of mass accretion rates, a key quantity for studying disk evolution. }
  % aims heading (mandatory)
   {Here we derive stellar properties of low-mass (\mstar$\lesssim$ 2 \msun) pre-main sequence stars without disks, which represent ideal photospheric templates for studies of young stars. We also use these spectra to constrain the impact of chromospheric emission on the measurements of mass accretion rates. The spectra are reduced, flux-calibrated, and corrected for telluric absorption, and are made available to the community.}
  % methods heading (mandatory)
   {We derive the spectral type for our targets by analyzing the photospheric molecular features present in their VLT/X-Shooter spectra by means of spectral indices and comparison of the relative strength of photospheric absorption features. We also measure effective temperature, gravity, projected rotational velocity, and radial velocity from our spectra by fitting them with synthetic spectra with the ROTFIT tool. The targets have negligible extinction ($A_V<0.5$ mag) and spectral type from G5 to K6, and from M6.5 to M8. They thus complement the library of photospheric templates presented by \citet{Manara13a}. We perform synthetic photometry on the spectra to derive the typical colors of young stars in different filters. We measure the luminosity of the emission lines present in the spectra and estimate the noise due to chromospheric emission in the measurements of accretion luminosity in accreting stars.  }
  % results heading (mandatory)
   {We provide a calibration of the photospheric colors of young pre-main sequence stars as a function of their spectral type in a set of standard broad-band optical and near-infrared filters. The logarithm of the noise on the accretion luminosity normalized to the stellar luminosity is roughly constant and equal to $\sim-$2.3 for targets with masses larger than 1 solar mass, and decreases with decreasing temperatures for lower-mass stars. For stars with masses of $\sim 1.5 M_\odot$ and ages of $\sim$1-5 Myr, the chromospheric noise converts to a limit of measurable mass accretion rates of $\sim 3\cdot 10^{-10} M_\odot$/yr. The limit on the mass accretion rate set by the chromospheric noise is of the order of the lowest measured values of mass accretion rates in Class II objects.
}
  % conclusions heading (optional), leave it empty if necessary 
   {}


   \maketitle
%
%________________________________________________________________

\section{Introduction}

The phase of pre-main sequence (PMS) stellar evolution is key for the final stellar mass build-up and for the evolution and dispersal of the surrounding circumstellar disks \citep[e.g.,][]{Hartmann16}. This, in turn, has a strong impact on  the final architecture of the planetary systems that form in the disk \citep{Morby16}. 

The determination of how the star accretes mass from the disk, how the system loses mass through winds, and how these processes affect disk evolution, requires a good knowledge of the properties of the central star. In particular, one needs to know its effective temperature ($T_{\rm eff}$) and luminosity (\lstar), and also the surface gravity (log$g$) and projected rotation velocity ($v$sin$i$), which are the main parameters shaping its photospheric emission flux. This photospheric flux must be subtracted from the observed spectrum in order to measure accretion and ejection diagnostics. The best approach seems to be to use as photospheric templates the spectra of disk-less non-accreting pre-main sequence stars, known as Class III objects based on their infrared classification, as they have similar surface gravity (intermediate between giants and dwarfs) to the accreting stars (Class II objects).  Current synthetic spectra for such low-surface-gravity objects are  not entirely reliable, as they do not fully reproduce  medium- or high-resolution spectra of low-gravity objects \citep[e.g.,][]{Allard11}. 
On the other hand, synthetic spectra are still valuable templates of non-active stars, and are therefore 
needed to evaluate the chromosperic flux emitted by very young stars. Therefore, spectra of Class~III objects are needed both as photospheric templates and to further calibrate synthetic spectra.

In \citet[][hereafter \citetalias{Manara13a}]{Manara13a} we presented  flux-calibrated spectra of 24 low-mass non-accreting PMS stars covering a large wavelength range ($\lambda\lambda\sim$ 300 - 2500 nm), obtained with the medium-resolution spectrograph VLT/X-Shooter.  This sample included objects with spectral type from K5 to M9.5 with typically two or more objects per spectral sub-class. The spectra, fully reduced and calibrated, were made available to the community and have already been used as templates for several studies \citep[e.g.,][]{Alcala14,Alcala17,Banzatti14,Fang16,Manara15,Manara16a,Manara17,Manjavacas14,Stelzer13b,Whelan14}. 
From the same spectra we estimated the contribution of  chromospheric activity to the emission lines typically used to estimate the accretion luminosity of accreting stars, showing that this has a strong dependence on the spectral type, and hence the stellar mass, of the target. A more detailed study of  the  photospheric and chromospheric properties of these stars, in particular their effective temperature, surface gravity, rotational velocity, and radial velocity, as well as the comparison between chromospheric and coronal emissions, was then carried out by \citet{Stelzer13}. This work showed that the chromospheric emission saturates for early-M stars and that the coronal flux dominates that of the chromosphere. 

Here we extend the library of photospheric templates with spectra of 16 additional non-accreting PMS stars presented here for the first time and one previously analyzed object. These additional objects have spectral types from G5 to M8, in particular in the ranges not covered by \citetalias{Manara13a}. The total sample of 41 reduced spectra, including one already analyzed in \citet{Manara16a}, is made available through the Centre de Donn\'ees astronomiques de Strasbourg (CDS) \footnote{http://vizier.u-strasbg.fr/viz-bin/VizieR}.
We present the sample selection, observations, and data reduction in Sect.~\ref{sect::obs}. Then, we discuss the analysis of the spectra and the photospheric properties of these objects in Sect.~\ref{sect::starprop}. In Sect.~\ref{sect::lines} we analyze the emission lines present in the spectra and the implications for measurements of mass accretion rates in PMS objects with disks. Finally, we perform synthetic photometry on the spectra and derive the typical colors for PMS stars in Sect.~\ref{sect::photcol}.

%__________________________________________________________________

\section{Sample, observations, and data reduction}\label{sect::obs}

The targets were selected to be bona-fide young stellar objects without a disk, that is, Class~III according to classifications based on Spitzer data \citep[e.g.,][]{Evans09}, and with negligible or very low extinction, that is, $A_V\lesssim$0.5 mag. Moreover, we searched for objects classified in the literature to have spectral type in the G and early K spectral classes, or in the late M spectral class, in order to cover the gaps in spectral types of the library of photospheric templates of young stars studied by \citetalias{Manara13a} and \citet{Stelzer13}. We based our selection on the works of \citet{Wahhaj10}, \citet{Manoj11}, \citet{Luhman07}, and \citet{Luhman08} and selected 16 single stars for observations. 
The objects are located in the Taurus, Chamaleon~I, Lupus, and Upper Scorpius star-forming regions. Coordinates, information from the literature and archival photometry for these targets are reported in Tables~\ref{tab::lit}-\ref{tab::phot} of the Appendix~\ref{app::lit}. Information on the binarity of the targets was searched for in the literature \citep[e.g.,][]{Nguyen12,Daemgen15} and is reported in Table~\ref{tab::lit}. 

Observations were carried out with the ESO VLT/X-Shooter spectrograph \citep{Vernet11}. This spectrograph covers simultaneously the wavelength range from $\sim$300 nm to $\sim$2500 nm, and the spectra are divided into three arms, the UVB ($\lambda\lambda\sim$ 300-550 nm), the VIS ($\lambda\lambda \sim$ 500-1050 nm), and the NIR ($\lambda\lambda\sim$ 1000-2500 nm). Slits with different widths were used in the three arms and for brighter or fainter targets. The brighter earlier-type objects were observed with the narrower slits (0.5\arcsec-0.4\arcsec-0.4\arcsec \ in the three arms, respectively) leading to the highest spectral resolution (R$\sim$9900, 18200, 10500 in the three arms, respectively). The targets with later spectral type were instead observed using larger slits (1.3\arcsec-0.9\arcsec-0.9\arcsec) leading to lower spectral resolution (R$\sim$ 4000, 7450, 5300) but also to lower flux losses. All the targets were also observed with wider slits of 5.0\arcsec \ prior to the narrow slit observations to obtain a spectrum with no flux losses needed for absolute flux calibration. The log of the observations is reported in Table~\ref{tab::log}. The S/N of the spectra depends on the wavelength and on the brightness of the targets. The targets with G- and K-spectral types (see following Section) have a mean S/N$>$60-100, with at least S/N$\sim$5-10 at 355 nm and increasing with wavelengths. The spectra of the four later-type objects have mean S/N$\sim$20-30, ranging from S/N$\sim$5-10 at $\lambda\sim$700 nm and smaller, even zero, at shorter wavelengths, to $\sim$40-60 in the NIR arm of the spectra.

The data reduction was carried out with the ESO X-Shooter pipeline v.2.5.2 \citep{Modigliani10}. This includes the usual reduction steps, such as flat fielding, bias subtraction, order extraction and combination, rectification, wavelength calibration, flux calibration using standard stars observed in the same night, and extraction of the spectrum. Two additional steps were performed by us on the pipeline-reduced spectra. First, removal of telluric absorption lines was performed using telluric standard stars observed close in time and airmass following the procedure described by \citet{Alcala14} both in the VIS arm, and also in the NIR arm. Differently from \citet{Alcala14}, here we also perform the telluric correction of the NIR spectrum using the flux calibrated spectrum from the pipeline. Then, the spectra obtained with narrow slits were corrected for slit losses by matching those to the one obtained using wider slits, thus obtaining absolute flux calibration. The correction factor is usually constant with wavelength for the objects observed with the 1.3\arcsec and 0.9\arcsec \ wide slits, while a correction factor with a linear dependence on wavelength was applied to the spectra obtained with the narrower slits. We finally checked all the flux-calibrated spectra and the agreement with archival photometry is excellent. The only peculiar case was RXJ0445.8+1556, for which an additional factor of $\sim$1.5 was needed to match the available photometry, probably due to a passage of a cloud while the exposure of the large slit was performed, making this spectrum dimmer than the one obtained with the narrower slits. The lithium absorption line at $\lambda\sim$670.8 nm is detected in all targets with G- or K-class spectral type, and also in LM717.

Additional X-Shooter spectra of Class~III objects were presented by \citet{Manara14} and \citet{Manara16a} and used in these works. However, all but one of these targets have $A_V>$ 1 mag, thus they are not included in this work, as we want to include only objects with negligible or very low extinction. We thus include in this work only HBC407, a K0 star with $A_V$ = 0.8 mag that was analyzed by \citet{Manara16a}.

All the 16 reduced, flux-calibrated, telluric corrected spectra analyzed here, plus the spectrum of HBC407 and the 24 targets analyzed by \citetalias{Manara13a} are available electronically on CDS.

%%%%%%%%%%%%%%%%%%%%%%%%%%%%%%%%%%%%%
\begin{figure}[!t]
\centering
\includegraphics[width=0.5\textwidth]{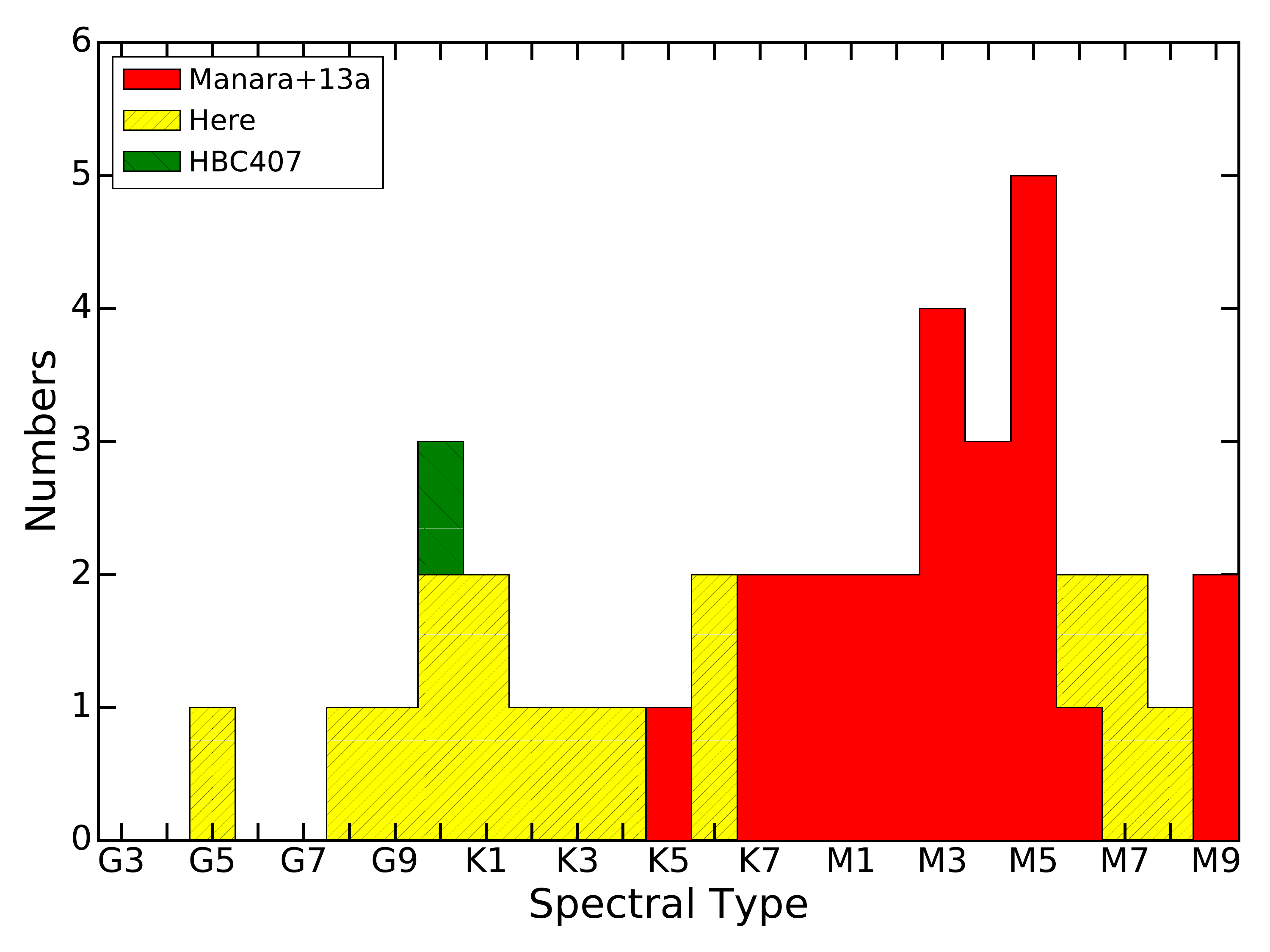}
\caption{Histogram of the spectral type of the photospheric templates observed with X-Shooter and analyzed here, by \citetalias{Manara13a}, and by \citet{Manara16a}.
     \label{fig::hist_spt}}
\end{figure}
%%%%%%%%%%%%%%%%%%%%%%%%%%%%%%%%%%%%%%%%%%%%%%%%%%%%%%%%%%%%%%%%%%%%%%%%%%%%

%__________________________________________________________________

\section{Stellar properties}\label{sect::starprop}

%%%%%%%%%%%%%%%%%%%%%%%%%%%%%%%%%%%%%%%%%%%%%%%%%%%%%%
% \setlength{\tabcolsep}{3pt}
% data of 2016-0-02 - added to the paper on 2017-09-01
\begin{table}
\begin{center}
\footnotesize
\caption{\label{tab::spt_ind} Spectral types obtained from spectral indices}
\begin{tabular}{l|c|ccc}  
\hline \hline
 Object &  \multicolumn{4}{c}{Spectral type}\\
\hbox{} & This work & Riddick & TiO & HH14 \\

\hline

RXJ0445.8+1556   &    G5       & \nodata & \nodata & \nodata \\
RXJ1526.0-4501      &  G9  & \nodata & \nodata & \nodata \\
PZ99J160550.5-253313 & K1      & \nodata &  K5.7 &    K0.8 \\
RXJ1508.6-4423    &   G8  & \nodata & \nodata & \nodata \\
PZ99J160843.4-260216 & K0.5     & \nodata & K5.9 &      K0.5 \\
RXJ1515.8-3331   &     K0.5  & \nodata  &   K5.6 &  K0.4     \\
RXJ1547.7-4018    &    K3    & \nodata &  K5.5 & K2.7 \\
RXJ0438.6+1546   &     K2  & \nodata &  K5.6 &    K1.5 \\
RXJ0457.5+2014   &     K1   &  \nodata &   K5.9 &   K1.2 \\
RXJ1538.6-3916    &    K4   & \nodata &    K5.7 & K4.5 \\
RXJ1540.7-3756    &    K6    &  \nodata &  K6.5 &    K9.5 \\
RXJ1543.1-3920    &    K6     & \nodata & K6.4 &  K8.9 \\
\hline
LM717     &            M6.5  & M6.4 &  M6.4 &   M6.7\\
LM601      &           M7.5  & M7.2 &  M7.7 &   M7.5   \\
J11195652-7504529  &   M7 &   M6.9 & M7.0 &   M7.3  \\
CHSM17173       &      M8   & M7.7 &  M7.3 &   M8.0 \\

\hline

\end{tabular}

\end{center}
\end{table}
%%%%%%%%%%%%%%%%%%%%%%%%%%%%%%%%%%%%%%%%%%%%%%%%%%%%%%

\subsection{Spectral type classification using spectral indices}\label{sect::spt}

Spectral types were derived using spectral indices and these estimates have been further confirmed by visually inter-comparing the depth of temperature-sensitive molecular features in the spectra analyzed here. The spectral indices used are the same as in \citetalias{Manara13a}, which were selected among the indices provided by \citet{Riddick07}, and augmented with other spectral indices by \citet[][hereafter \citetalias{HH14}]{HH14} and the TiO index by \citet{Jeffries07}. These additional indices are particularly useful to assign a spectral type to stars of K- and early M- spectral type, which correspond to a range of stellar temperatures to which the \citet{Riddick07} indices, developed for later-type objects, are not sensitive. The TiO index by \citet{Jeffries07}, in particular, is sensitive to spectral types K5, or later. 
The spectral type was thus assigned using the indices by \citetalias{HH14} for stars with spectral type between K0 and K5, the TiO index for stars with K6 and later K spectral type, and the mean values of the three sets of indices for the M spectral type objects. All values were rounded to half a subclass. The values derived with the different indices are reported in Table~\ref{tab::spt_ind}. The spectra of the objects are shown, ordered by spectral type, in Fig.~\ref{fig::early1}-\ref{fig::late}.

%%%%%%%%%%%%%%%%%%%%%%%%%%%%%%%%%%%%%%%%%%%%%%%%%%%%%%
% \setlength{\tabcolsep}{3pt}
% version of 2015-01-05
\begin{table*}
\begin{center}
\footnotesize
\caption{\label{tab::star_pars} Stellar parameters derived for the targets analyzed in this work }
\begin{tabular}{l|cc|cccc|cc}  
\hline \hline
 Object &  RA(2000)  & DEC(2000) & $d$ & SpT & T$_{\rm eff}$ &  log($L_\star/L_\odot$) & $d_{\rm TGAS}$ & log($L_\star/L_\odot$)$_{\rm TGAS}$  \\       &  h \, :m \, :s & $^\circ$ \, ' \, ''   &  [pc] & \hbox{} & [K]  & & [pc] & \\         
\hline

RXJ0445.8+1556  &               04:45:51.29 &+15:55:49.69       & 140 &G5       &       5770    &       0.485  & 141$\pm$4 & 0.497 \\
RXJ1526.0-4501  &               15:25:59.65 &$-$45:01:15.72     & 150 & G9      & 5410 & $-$0.061 & 147$\pm$5 & $-$0.073 \\
PZ99J160550.5-253313    &       16:05:50.64& $-$25:33:13.60 & 145 & K1  & 5000    & $-$0.007  & 105$\pm$6 & $-$0.287\\
RXJ1508.6-4423  &               15:08:37.75 & $-$44:23:16.95&   150 & G8        & 5520    & 0.043 & 148$\pm$12 & 0.031 \\
PZ99J160843.4-260216            &16:08:43.41 &$-$26:02:16.84    & 145 & K0.5    & 5050 & 0.140  & 152$\pm$12 & 0.181\\
RXJ1515.8-3331          &       15:15:45.36 &$-$33:31:59.78&    150 & K0.5      & 5050    & 0.098 & \nodata & \nodata \\
RXJ1547.7-4018  &               15:47:41.76 &$-$40:18:26.80&    150 & K3        & 4730 & $-$0.081 & 135$\pm$4 & $-$0.173 \\
RXJ0438.6+1546          &       04:38:39.07& +15:46:13.61&      140 &K2 & 4900    & $-$0.024  & 145$\pm$5 & 0.006 \\
RXJ0457.5+2014          &       04:57:30.66 &+20:14:29.42&      140 & K1        & 5000    & $-$0.150 & \nodata & \nodata \\
RXJ1538.6-3916          &       15:38:38.36 &$-$39:16:54.08&    150 & K4        & 4590 & $-$0.217 & \nodata & \nodata \\
RXJ1540.7-3756          &       15:40:41.17& $-$37:56:18.54&    150 & K6        & 4205    & $-$0.405 & \nodata & \nodata \\
RXJ1543.1-3920          &       15:43:06.25 & $-$39:20:19.5&    150 & K6 &       4205 &  $-$0.397 & \nodata & \nodata \\
\hline
LM717           &               11:08:02.34 &$-$76:40:34.3              & 160 &M6.5       & 2935 &        $-$1.750 & \nodata & \nodata \\
LM601                   &       11:12:30.99& $-$76:53:34.2              &160 & M7.5  & 2795 & $-$2.229 & \nodata & \nodata \\
J11195652-7504529       &       11:19:56.52 &$-$75:04:52.9      &160 &  M7      & 2880 & $-$2.177 & \nodata & \nodata \\
CHSM17173               &       11:10:22.26 &$-$76:25:13.8              &160 &M8     & 2710 & $-$1.993 & \nodata & \nodata  \\

\hline

\end{tabular}

\end{center}
\end{table*}
%%%%%%%%%%%%%%%%%%%%%%%%%%%%%%%%%%%%%%%%%%%%%%%%%%%%%%

For the classification of the objects with G spectral type we used a method introduced by \citetalias{HH14}. We join with a straight line the continuum emission at $\lambda\lambda$ 460 nm and 540 nm and   between $\lambda\lambda$ 490 nm and 515 nm. We then compute the distance between these two lines at  $\lambda$ = 515 nm. 
This distance increases with  the  spectral type  for G  and K stars,  as  shown in Fig.~\ref{fig::early2}. This method allows us to classify G-type objects and, in turn, confirms the classification for K-type objects. 

The final values of spectral type for the targets, and the corresponding effective temperatures, are reported in Table~\ref{tab::star_pars}. The uncertainties on the spectral type are typically one sub-class for objects of G and K spectral types in our sample, and half a sub-class for M-type objects. Our estimates are typically consistent within one spectral sub-class with those from the literature (cf. Table~\ref{tab::lit}), and different by two sub-classes at most. We show in Fig.~\ref{fig::hist_spt} the histogram of the spectral types of the targets discussed here together with those analyzed by \citetalias{Manara13a}. When considered together, these libraries are now covering the whole range of spectral type from G8 to M9.5 with at least one template per spectral sub-class.

\subsection{Extinction estimate}

The targets analyzed here were selected to have negligible extinction ($A_V<$0.5 mag). We verify this assumption with two methods. 
%%%%%%%%%%%%%%%%%%%%%%%%%%%%%%%%%%%%%
\begin{figure*}[]
\centering
\includegraphics[width=\textwidth]{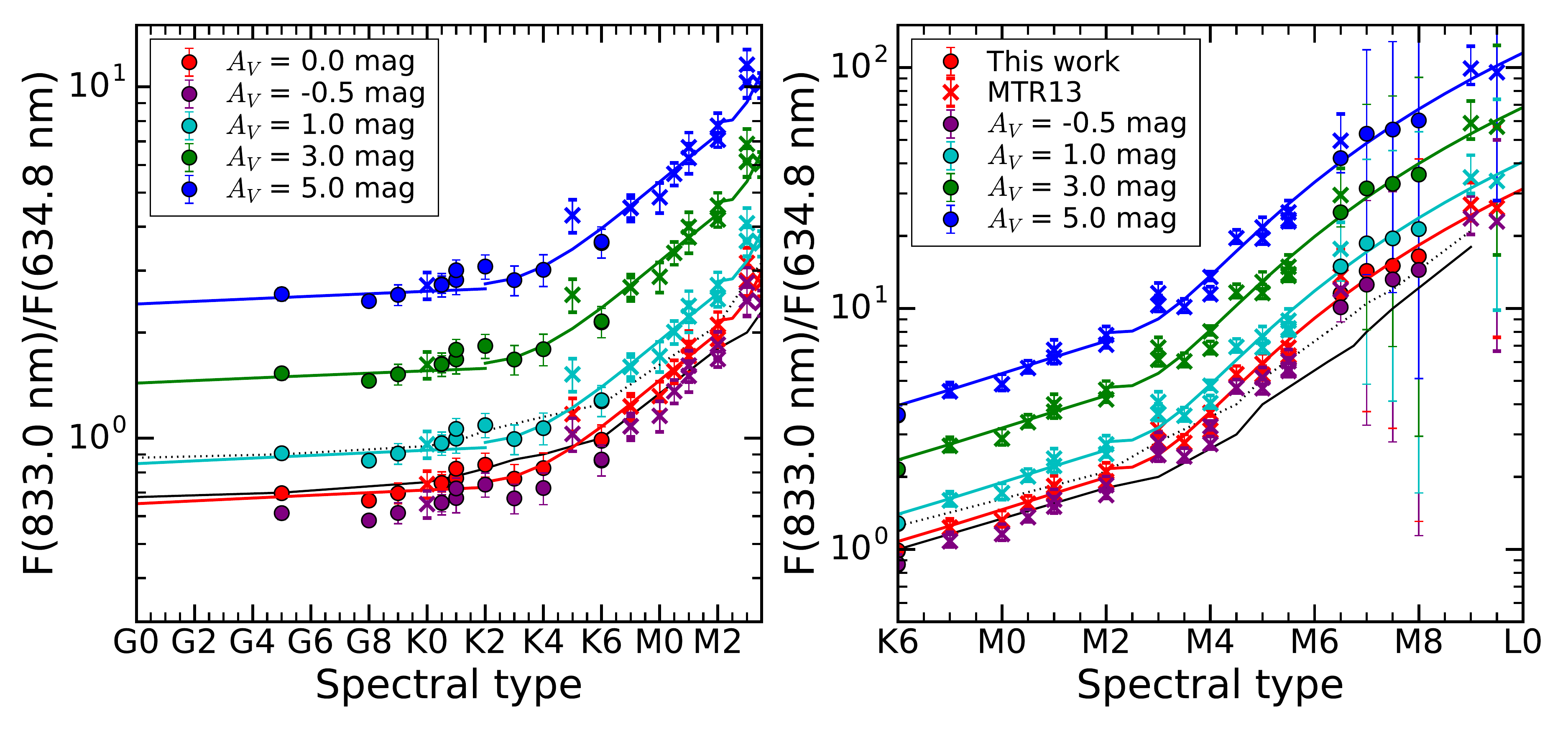}
\caption{$F_{\rm red} = F(833.0 {\rm nm})/F(634.8 {\rm nm})$ ratio calculated for the ClassIII spectra analyzed here (circles) and by \citetalias{Manara13a} (crosses). Different colors refer to values of $F_{\rm red}$ calculated on spectra reddened with increasing values of $A_V$, from 0 mag to 5 mag, as reported in the legend. The data reported as $A_V$ = -0.5 mag are obtained by de-reddening the spectrum by $A_V$ = 0.5 mag. The black solid line is the value of  $F_{\rm red}$ derived by \citetalias{HH14} for $A_V$ = 0 mag, and the dotted line for $A_V$ = 1 mag. The colored lines are our new estimates of this ratio, which are reported in Table~\ref{tab::flred}.
     \label{fig::flred}}
\end{figure*}
%%%%%%%%%%%%%%%%%%%%%%%%%%%%%%%%%%%%%%%%%%%%%%%%%%%%%%%%%%%%%%%%%%%%%%%%%%%%

First, we compare the observed spectra to BT-Settl models \citep{Allard11} with the same temperature as the target and with log$g$=4.0 reddened using the extinction law by \citet{Cardelli} and $R_V$=3.1 with increasing values of $A_V$ in steps of 0.1 mag, and normalized at 750 nm to the continuum of the observed spectra. The value of $A_V$ at which the squared difference between the target and the model is minimized is then chosen. This method, when applied to G- and K-type objects, leads to values of $A_V$=0 mag for RXJ1508.6-4423, RXJ1547.7-4018, RXJ1538.6-3916, and RXJ1540.7-3756, of $A_V$=0.1 mag for RXJ1526.0-4501 and RXJ1543.1-3920, of $A_V$=0.2 mag for PZ99J160550.5-253313, PZ99J160843.4-260216, RXJ1515.8-3331, and RXJ0457.5+2014, of $A_V$=0.3 mag for RXJ0438.6+1546 and $A_V$=0.4 mag for RXJ0445.8+1556. These values have typically 1$\sigma$ uncertainties of 0.1 mag, and of 0.2 mag for the highest values. The values of extinction are thus, in all cases, smaller than 0.5 mag, and compatible with $A_V$=0 mag within, at most, 3$\sigma$. In the case of the four late-M-type objects, the agreement between the synthetic spectra and the observed one is not good. In particular, it is not possible to have a good agreement in both the near-infrared part of the spectra ($\lambda > 1000$ nm) and the optical part ($\lambda>500$ nm) simultaneously. To check the values of $A_V$ we use a normalization at 1040 nm, which allows us to find a better match in the near-infrared part of the spectrum. We obtain $A_V$=1.3, 0.3, 0.9, and 0.6 mag for LM717, LM601, J11195652-7504529, and CHSM17173, respectively. However, these values are rather uncertain and are highly susceptible to the choice of the normalization wavelength, of the reddening law, and to the uncertainties in the models. We note that the values of $A_V$ estimated from the near-infrared colors for these late-type targets by \citet{Luhman07} are of 0 mag for all the objects apart from LM717, which was estimated at $A_V$=0.4 mag (see Table~\ref{tab::lit}).  Differences in the values of $A_V$ derived from near-infrared or optical spectra could be due to several physical properties of the objects, such as the presence of cold spots on the stellar surface \citep[e.g.,][]{Stauffer03,Vacca11,Pecaut16,Gully-Santiago17}, which make the targets appear to be of later types in red spectra than in blue spectra.

An independent estimate of $A_V$ can be obtained from the ratio $F_{\rm red} = F(833.0 {\rm nm})/F(634.8 {\rm nm})$, as first suggested by \citetalias{HH14}. This  ratio is  sensitive to $A_V$ and depends on the spectral type of the object. Fig.~\ref{fig::flred} shows the observed values  for the objects analyzed here together with those in \citetalias{Manara13a} \footnote{The spectrum of HBC407 used here is de-reddened by $A_V$ = 0.8 mag, and is considered on the plot together with those of \citetalias{Manara13a}.}, as well as  the values obtained after applying a reddening correction with increasing values of $A_V$ from 0 mag to 5 mag, using the reddening law by \citet{Cardelli} and $R_V$=3.1. We also show the value of $F_{\rm red}$  after de-reddening the spectra by $A_V$ = 0.5 mag, reported on the plot as a negative value of $A_V$. On the same figure we show the values of this ratio calculated by \citetalias{HH14} for $A_V$ = 0 mag and 1 mag. The observed values of $F_{\rm red}$ are consistent with our assumption that $A_V\sim 0$ mag for spectral types earlier than M. As for the comparison with synthetic spectra, the values of $A_V$ for objects with G- and K spectral type are negligible.
There are, however, differences in the M sub-class range. This could be due to the larger uncertainty in the estimate of this ratio in the analysis by \citetalias{HH14}, who used both synthetic spectra and observed spectra of dwarfs to calibrate the index. Since synthetic spectra are not reliable in this spectral-type range \citep[e.g.,][]{DaRio10,Bell12}, we argue that all the spectra of our targets are compatible with having negligible extinction, that is, $A_V<$ 0.5 mag, and we use our data to re-calibrate the values of $F_{\rm red}$ as a function of spectral type and $A_V$. The new values of $F_{\rm red}$ are reported in Table~\ref{tab::flred}.

%%%%%%%%%%%%%%%%%%%%%%%%%%%%%%%%%%%%%
\begin{figure}[!t]
\centering
\includegraphics[width=0.5\textwidth]{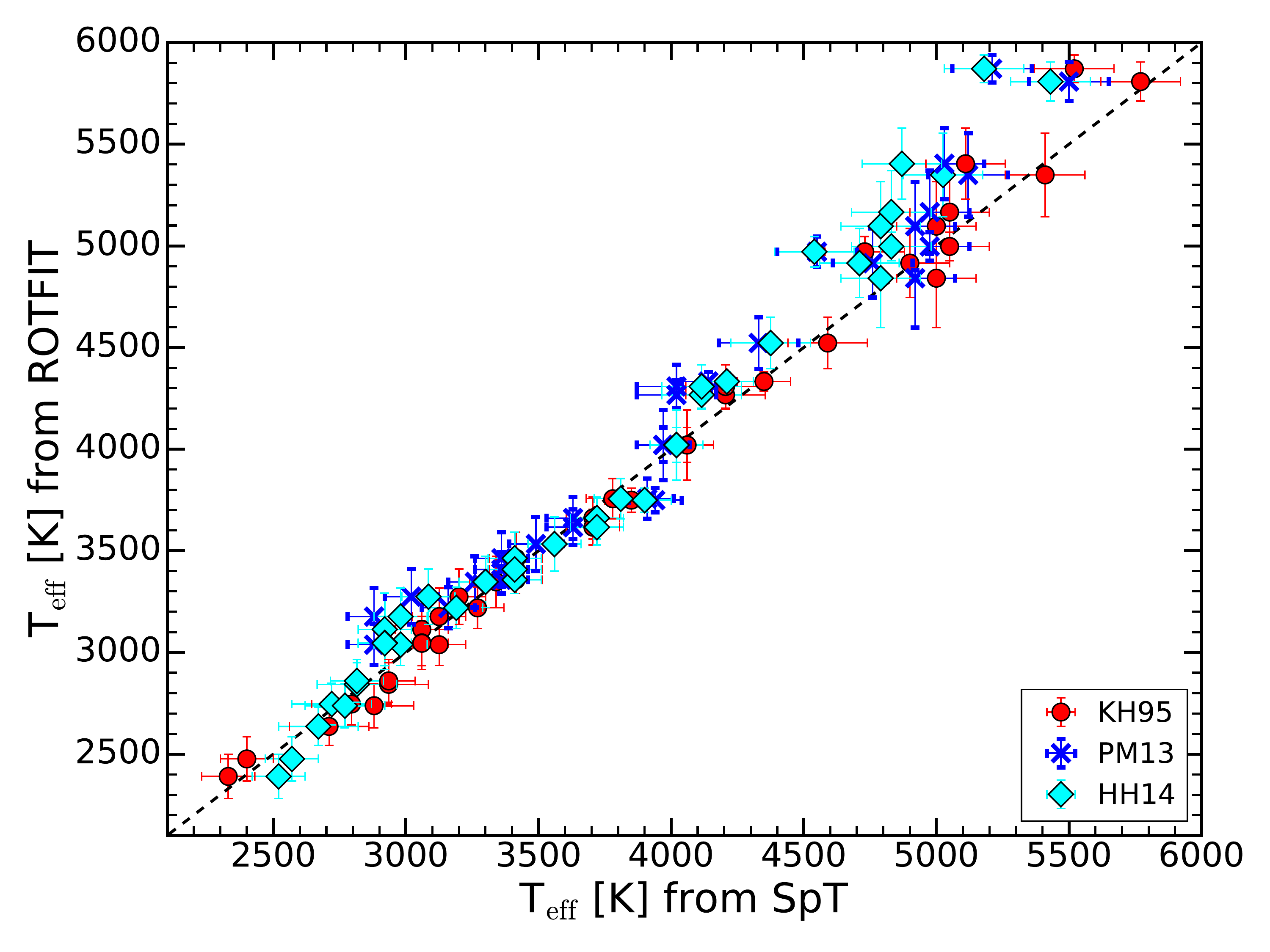}
\caption{Comparison between temperatures from spectral types and from ROTFIT for the objects analyzed by \citetalias{Manara13a}, who reported the spectral type for these targets, and by \citet{Stelzer13} who presented the ROTFIT analysis, and for those analyzed here. The relations between spectral type and temperature are those used by \citet{Manara13a}, labeled as `KH95' as it is mainly based on \citet{KH95}, as discussed in the text, the one by \citet{PM13}, labeled as `PM13', and the one by \citet{HH14}, labeled as `HH14'. The relation by \citet{PM13} is not available for spectral types later than M5.
     \label{fig::rotfit_cfm}}
\end{figure}
%%%%%%%%%%%%%%%%%%%%%%%%%%%%%%%%%%%%%%%%%%%%%%%%%%%%%%%%%%%%%%%%%%%%%%%%%%%%

%%%%%%%%%%%%%%%%%%%%%%%%%%%%%%%%%%%%%
\begin{figure}[!t]
\centering
\includegraphics[width=0.5\textwidth]{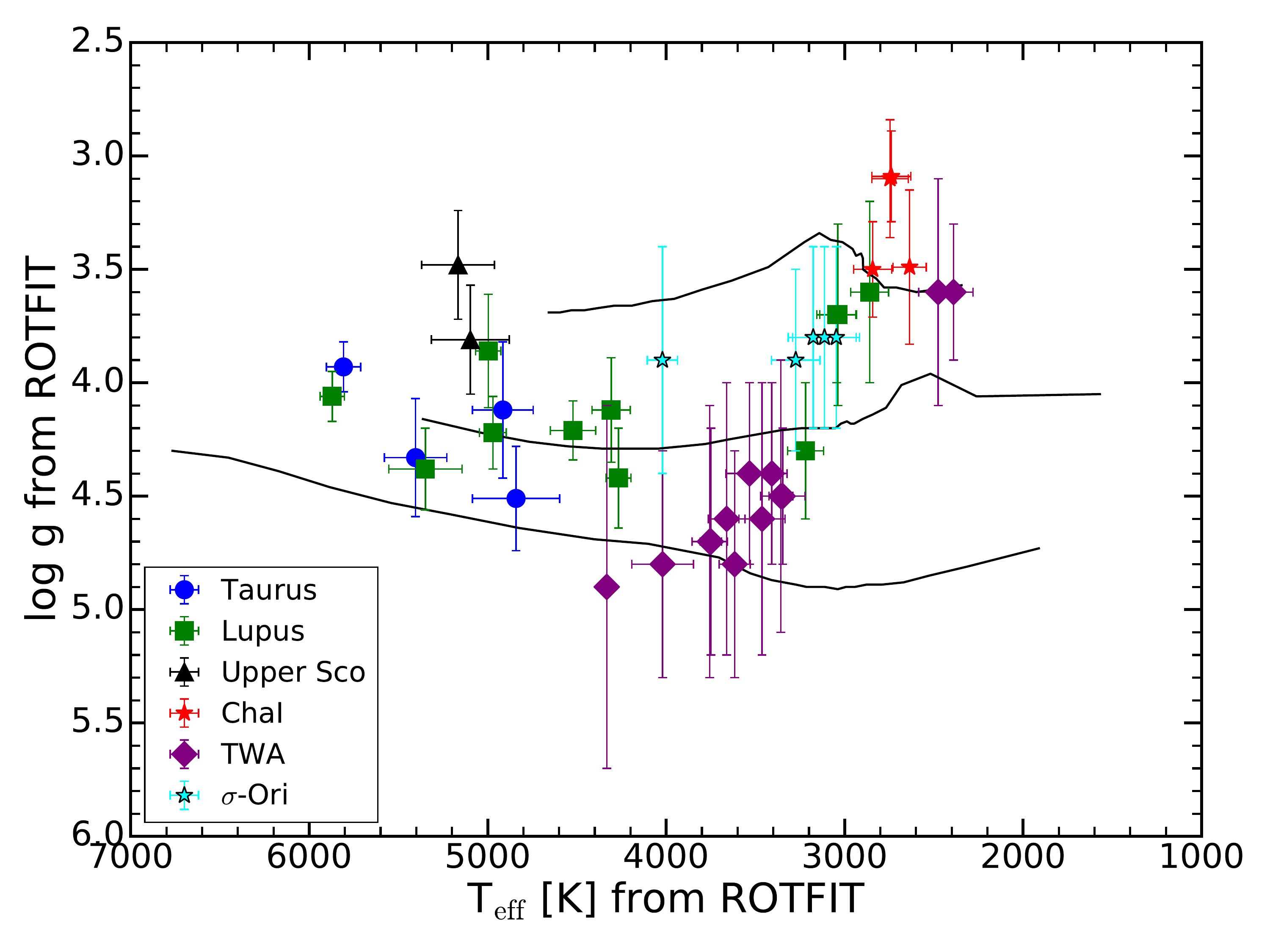}
\caption{\logg--\teff\  diagram for the targets analyzed here and by \citet{Stelzer13} showing the values derived with ROTFIT. Isochrones for 1, 10, and 100 Myr (from top to bottom) are from \citet{Baraffe15}.
     \label{fig::rotfit_logg_teff}}
\end{figure}
%%%%%%%%%%%%%%%%%%%%%%%%%%%%%%%%%%%%%%%%%%%%%%%%%%%%%%%%%%%%%%%%%%%%%%%%%%%%

\subsection{Photospheric properties from ROTFIT}\label{sect::rotfit}

The X-Shooter spectra have sufficient spectral resolution and wavelength coverage to  measure photospheric properties, namely the effective temperature (\teff), the surface gravity (\logg), the projected rotational velocity (\vsini), and  the radial velocity (RV) of the targets. We perform this analysis on the objects of our sample, including HBC407, using the ROTFIT tool \citep[e.g.,][]{Frasca15,Frasca17}, which was also applied by \citet{Stelzer13} to analyze the Class~III spectra of \citetalias{Manara13a}. This analysis tool is based on the search for the best template spectrum that reproduces the target spectrum by minimizing the $\chi^2$ of the difference between observed and template in specific spectral segments. We adopted as templates a grid of BT-Settl spectra \citep{Allard11} with solar iron abundance according to the \citet{Asplund09} solar mixture and effective temperature in the range 2000--6000\,K and \logg\  from 5.5 to 0.5 dex.
The tool is run setting  the veiling to  zero in all targets, since they are non-accreting objects. For the G- and K-type objects both the VIS and UVB arms have been used in the analysis with ROTFIT, while the VIS arm contains sufficient lines to analyze the colder objects in the sample. For the latter objects, only the VIS arm is used because of the lower S/N of the spectra in the UVB arm. The spectral intervals analyzed with ROTFIT contain features that are sensitive to the effective temperature and/or \logg, such as the \ion{Na}{i} doublet at $\lambda\approx 8190$\,\AA\  
and the \ion{K}{i} doublet at $\lambda\approx 7660-7700$\,\AA. The code
also allows us to measure the \vsini\  by $\chi^2$ minimization applied to spectral segments devoid of broad lines. The photospheric parameters derived with ROTFIT are reported in Table~\ref{tab::rotfit}. For completeness, we include also the values of \vsini\  and RV, even if they are not used in the remainder of this paper. 

Fig.~\ref{fig::rotfit_cfm} shows the comparison between $T_{\rm eff}$ derived with ROTFIT and the one obtained by converting the spectral type determined as in Sect.~\ref{sect::spt} using different relations between spectral type and $T_{\rm eff}$. We use here the relation by \citet{KH95} for objects with spectral type in the G and K classes, and by \citet{Luhman03} for later-type objects, in line with the analysis by \citetalias{Manara13a} and \citet{Stelzer13}, the one by \citet{PM13} for 5-30 Myr-old objects, and the one by \citetalias{HH14} for young stars.
The values obtained with these two methods using the combined relations of \citet{KH95} and \citet{Luhman03} agree within the 1$\sigma$ uncertainties for all targets except RXJ1508.6-4423, HBC407, and RXJ1547.7-4018. Other relations result in larger deviations especially for the stars with earliest types, which are, however, still within 3$\sigma$.

Fig.~\ref{fig::rotfit_logg_teff} shows the log$g$-$T_{\rm eff}$ plane for all targets in this work and those analyzed by \citet{Stelzer13}, overplotted with theoretical isochrones by \citet{Baraffe15}. All the targets are compatible within the uncertainties of the parameters with being young objects with ages $\lesssim$ 10 Myr. These values of log$g$ derived here are thus compatible with the fact that these objects are PMS, and thus closer to luminosity class IV. We note that the coldest stars, thus with lower masses, have smaller values of log$g$, as noted also by \citet{Stelzer13}. This would suggest a younger age for the objects, and we come back to this point in Sect.~\ref{sect::hrd}.

%%%%%%%%%%%%%%%%%%%%%%%%%%%%%%%%%%%%%%%%%%%%%%%%%%%%%%
% \setlength{\tabcolsep}{3pt}
\begin{table*}
\begin{center}
\footnotesize
\caption{\label{tab::rotfit} Photospheric parameters derived with ROTFIT }
\begin{tabular}{l|cc|cc|cc|cc}%|cc}  
\hline \hline

Object & $T_{\rm eff}$ &   $\sigma(T_{\rm eff})$ &   log$g$ &   $\sigma$(log$g$)  &  vsin$i$ &   $\sigma$(vsin$i$) &  RV  &   $\sigma$(RV) \\%&    RV$_{\rm UVB}$  &   $\sigma$(RV$_{\rm UVB}$ )  \\
\hbox{} & [K] & [K] & [cgs] & [cgs] & [km/s] & [km/s] & [km/s] & [km/s] \\%& [km/s] & [km/s]  \\

\hline

    RXJ0445.8+1556 & 5808     &   96     & 3.93  &    0.11 &    117.1  &     7.8    &  17.6     &  7.4 \\% &    \nodata   & \nodata  \\
    RXJ1526.0-4501 &    5349   &    205  &    4.38     & 0.18    &  25.0    &   1.0    &   5.7     &  2.0 \\%   &   6.7     &  2.9  \\
 PZ99J160550.5-253313 &     5097   &    218    &  3.81   &   0.24    &  13.0   &    6.0    &  -1.5   &    1.7  \\% &   10.0     &  2.8  \\
    RXJ1508.6-4423 &  5871    &    68  &    4.06   &   0.11   &  130,2    &   9.2   &    7.9    &  10.1\\% &     \nodata  & \nodata  \\
 PZ99J160843.4-260216 &   5166     &  204   &   3.48 &     0.24   &   42.5    &   1.1  &    -4.3   &    2.5   \\% &  -8.4    &   3.4  \\
    RXJ1515.8-3331 &  4997   &     71  &    3.86   &   0.25   &   22.3   &    1.0    &   2.0   &    1.9   \\% &   0.6   &    2.9  \\
    RXJ1547.7-4018 &     4971   &     75   &   4.22   &   0.16  &    11.1  &     1.0    &   2.9    &   1.7    \\%&   2.2     &  3.0    \\
    RXJ0438.6+1546 &     4915   &    170  &    4.12 &     0.30     & 26.3    &   1.0    &  18.6  &     1.9   \\%  & 19.5 &      3.0  \\
    RXJ0457.5+2014 &    4841   &    244    &  4.51  &    0.23     & 35.6    &   1.2    &  13.4    &   2.1   \\%  & 17.2    &   3.3  \\
    RXJ1538.6-3916 &   4522      & 127  &    4.21    &  0.13   &    1.0     &  2.0     &  2.6    &   1.7  \\% &    2.6  &     3.3    \\
    RXJ1540.7-3756 &     4267     &   70    &  4.42  &    0.22   &   19.1 &      1.0     &  1.7      & 1.9  \\%  &   1.9    &   3.6  \\
    RXJ1543.1-3920 &   4308   &    107  &    4.12  &    0.23    &  12.1     &  1.0     &  4.4     &  1.8   \\%   & 4.9    &   3.5  \\
\hline
                 LM717 &    2843      & 107     & 3.50  &    0.21    &  19.0  &    21.0   &   23.2   &    2.5  \\%  &   \nodata    &   \nodata  \\                 LM601 &   2746     &  102   &   3.10  &    0.26  &    19.0   &   26.0    &  20.3   &   11.1   \\%  &  \nodata   &    \nodata  \\
     J11195652-7504529 &    2738   &    109    &  3.09    &  0.20   &   44.0    &  19.0    &  15.1   &    5.6 \\%  &    \nodata    &   \nodata  \\             CHSM17173 &     2636    &    93   &   3.49  &    0.34    &  19.0   &   16.0   &   15.8    &   2.7   \\% &   \nodata    &   \nodata   \\
\hline

 HBC407 &    5404   &    175   &   4.33    &  0.26   &   10.0    &   1.0    &  19.1    &   1.7  \\%  &  18.8   &    2.6  \\

\hline

\end{tabular}

\end{center}
\end{table*}
%%%%%%%%%%%%%%%%%%%%%%%%%%%%%%%%%%%%%%%%%%%%%%%%%%%%%%

\subsection{Stellar luminosity}\label{sect::hrd}

The stellar luminosities are obtained with the same procedure as in \citetalias{Manara13a}. The flux of the observed X-Shooter spectrum is integrated over the whole wavelength range, apart from the telluric bands in the NIR arm, where the continuum of the spectrum is interpolated with a straight line. The flux from the regions shortwards of $\sim$300 nm and at wavelengths longer than $\sim$2500 nm, which are not covered by X-Shooter, is obtained by integrating a BT-Settl synthetic spectrum \citet{Allard11} with the same $T_{\rm eff}$ as the target and with log$g$ = 4.0, after matching it with the observed spectrum. We check that this method leads to results compatible with those obtained using the bolometric correction by \citetalias{HH14}. The stellar luminosity (\lstar) is then derived from the bolometric flux using the distances to the star-forming regions where the targets are located, namely 140 pc for Taurus \citepalias{HH14}, 150 pc for Lupus \citep{Comeron08}, 145 pc for Upper Sco \citep{deZeeuw}, and 160 pc for Chamaeleon~I \citep{LuhmanCha}. The computed stellar luminosities  are reported in Table~\ref{tab::star_pars}. We also report in the last two columns of Table~\ref{tab::star_pars} the recent distance estimates from the TGAS catalog contained in the Data Release 1 \citep{GaiaDR1} of the Gaia satellite \citep{Gaiamission}\footnote{The conversion from parallax to distance is taken from \citet{Astraatmadja16} and assumes the Milky way prior. The quoted errors do not include the systematic uncertainty of 0.3 mas.}, as well as the stellar luminosity corresponding to these revised distances. Distances obtained from parallax measurements are available for seven targets, with only the distance to PZ99J160550.5-253313 which significantly differs from the ones usually assumed for these objects. The revised distance to this object would change its age from $\sim$15 Myr to $\sim$30 Myr according to the evolutionary models by \citet{Siess00}.

\begin{figure}[!t]
\centering
\includegraphics[width=0.5\textwidth]{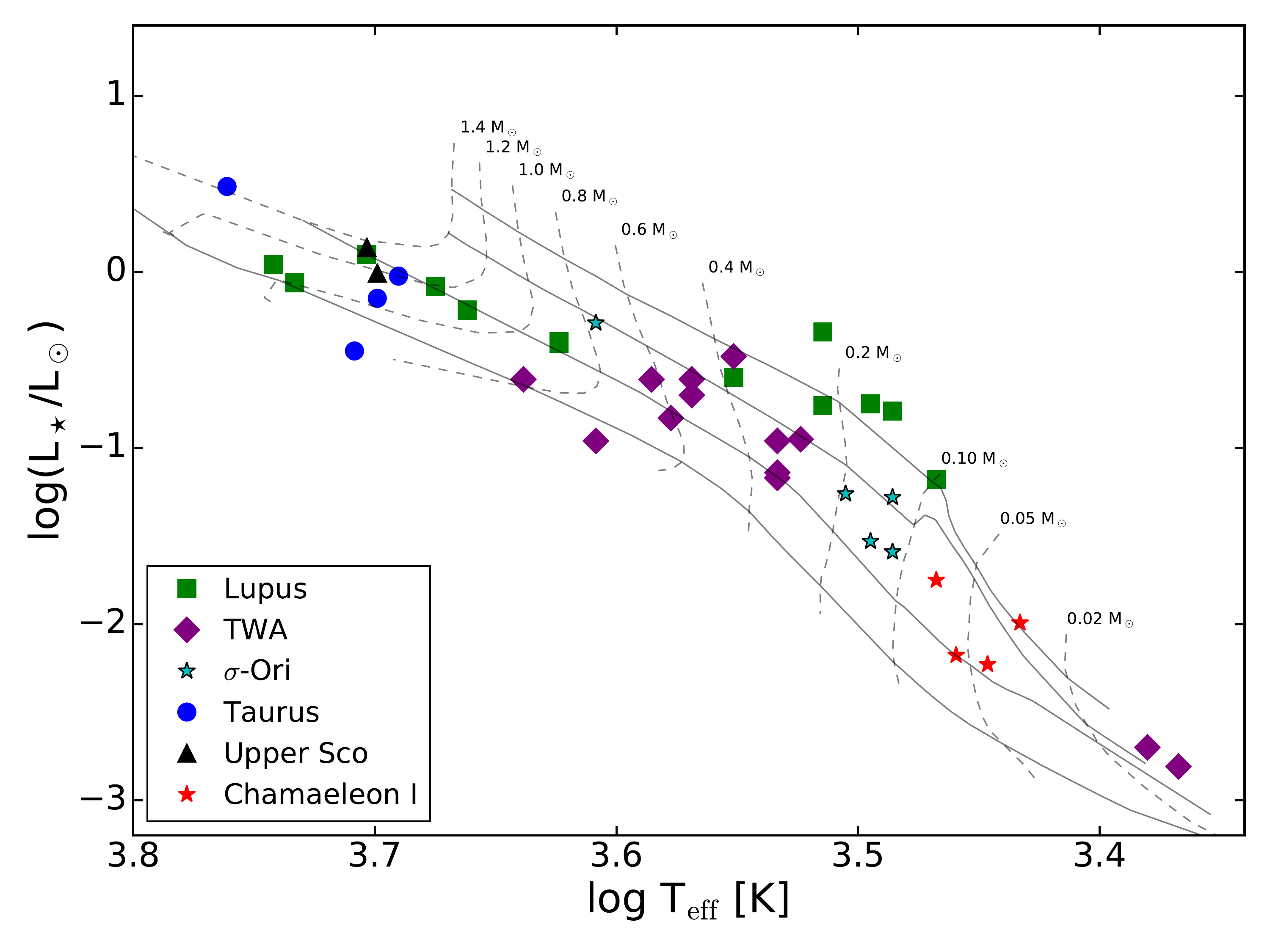}
\caption{HR diagram of the objects analyzed here and those by \citetalias{Manara13a}, where $T_{\rm eff}$ is obtained from the spectral type. The models by \citet{Baraffe15} are also shown. The isochrones are the 1.2, 3, 10, and 30 Myr ones. 
     \label{fig::HRD1}}
\end{figure}

The Hertzsprung-Russel diagram (HRD) is shown in Fig.~\ref{fig::HRD1}, with the evolutionary tracks of  \citet{Baraffe15} superimposed. This diagram shows an apparent mass-dependent isochronal age trend, with lower-mass stars being typically $\sim$1-10 Myr old, and stars with \mstar$\gtrsim$0.8 \msun \ being $\sim$10-30 Myr old. This trend is observed even in objects belonging to the same star-forming region, for example, Lupus, and is in line with other observations in different star-forming regions \citep[e.g.,][]{Bell14,HH15,PM16}, and with the log$g$-$T_{\rm eff}$ trend we see in Fig.~\ref{fig::rotfit_logg_teff}. Possible origins for this dependence are the presence of spots on the stellar surface, or the effects of convection related to the presence of magnetic fields, or the effects of accretion during the pre-main sequence evolutionary phase, which are currently being implemented in evolutionary models \citep[e.g.,][]{Somers15,Feiden16,Baraffe17}. It is also worth noting that the ages inferred from the log$g$-$T_{\rm eff}$ relation differ from those inferred from the HRD. For example, the targets in the Chamaeleon~I region are always younger than 1 Myr in the former, and have age $\sim$1-10 Myr in the latter diagram. This discrepancy is in line with previous results \citep[e.g.,][]{Slesnick06}.

%%%%%%%%%%%%%%%%%%%%%%%%%%%%%%%%%%%%%
%%%%%%%%%%%%%%%%%%%%%%%%%%%%%%%%%%%%%%%%%%%%%%%%%%%%%%%%%%%%%%%%%%%%%%%%%%%%

%%%%%%%%%%%%%%%%%%%%%%%%%%%%%%%%%%%%%%%%%%%%%%%%%%%%%%
\begin{table*}
\begin{center}
\footnotesize
\caption{\label{tab::phot_col} Photospheric colors of young non-accreting stars}
\begin{tabular}{l|cccccccc|cccccc}  
\hline \hline

Spectral & \multicolumn{2}{c}{$U-B$}  &      \multicolumn{2}{c}{$B-V$}    &     \multicolumn{2}{c}{$V-R$}    &     \multicolumn{2}{c}{$R-I$}  & \multicolumn{2}{c}{$V-J$}   & \multicolumn{2}{c}{$J-H$}    &     \multicolumn{2}{c}{$H-Ks$}    \\
Type & [mag] & [mag]& [mag]& [mag]& [mag]& [mag] & [mag] & [mag] & [mag]& [mag]& [mag]& [mag]& [mag] & [mag] \\

\hline

  G5   &   0.30   & $^{+0.94}_{-0.45}$ &    0.72   & $^{+0.24}_{-0.22}$  &    0.44  & $^{+0.35}_{-0.19}$ &     0.43   & $^{+0.15}_{-0.12}$  &     1.46  & $^{+1.18}_{-0.50}$ & 0.42   & $^{+0.17}_{-0.21}$  &   0.00    & $^{+0.15}_{-0.18}$ \\[3pt]
  G6   &   0.25   & $^{+0.71}_{-0.32}$ &    0.71   & $^{+0.22}_{-0.15}$  &    0.42  & $^{+0.29}_{-0.15}$ &     0.41   & $^{+0.13}_{-0.10}$  &     1.40  & $^{+0.83}_{-0.38}$ & 0.41   & $^{+0.16}_{-0.14}$  &   0.04    & $^{+0.11}_{-0.16}$ \\[3pt]
  G7   &   0.25   & $^{+0.55}_{-0.24}$ &    0.73   & $^{+0.18}_{-0.11}$  &    0.42  & $^{+0.23}_{-0.12}$ &     0.41   & $^{+0.11}_{-0.07}$  &     1.40  & $^{+0.71}_{-0.30}$ & 0.42   & $^{+0.14}_{-0.10}$  &   0.06    & $^{+0.09}_{-0.14}$ \\[3pt]
  G8   &   0.29   & $^{+0.46}_{-0.19}$ &    0.77   & $^{+0.14}_{-0.10}$  &    0.44  & $^{+0.17}_{-0.10}$ &     0.42   & $^{+0.08}_{-0.06}$  &     1.44  & $^{+0.58}_{-0.29}$ & 0.44   & $^{+0.12}_{-0.09}$  &   0.08    & $^{+0.08}_{-0.12}$ \\[3pt]
  G9   &    0.35  & $^{+0.35}_{-0.19}$  &    0.82  & $^{+0.12}_{-0.09}$   &    0.46 & $^{+0.13}_{-0.08}$  &     0.44  & $^{+0.07}_{-0.05}$   &    1.51  & $^{+0.46}_{-0.25}$ &  0.47  & $^{+0.10}_{-0.08}$   &   0.08   & $^{+0.08}_{-0.10}$  \\[3pt]
  K0   &    0.44  & $^{+0.29}_{-0.17}$  &    0.87  & $^{+0.09}_{-0.08}$   &    0.49 & $^{+0.10}_{-0.08}$  &     0.46  & $^{+0.06}_{-0.05}$   &    1.60  & $^{+0.41}_{-0.23}$ &  0.50  & $^{+0.09}_{-0.07}$   &   0.09   & $^{+0.08}_{-0.08}$  \\[3pt]
  K1   &    0.54  & $^{+0.26}_{-0.16}$  &    0.93  & $^{+0.09}_{-0.09}$   &    0.52 & $^{+0.10}_{-0.08}$  &     0.49  & $^{+0.05}_{-0.06}$   &    1.72  & $^{+0.38}_{-0.25}$ &  0.53  & $^{+0.08}_{-0.07}$   &   0.10   & $^{+0.08}_{-0.07}$  \\[3pt]
  K2   &    0.65  & $^{+0.27}_{-0.16}$  &    0.99  & $^{+0.10}_{-0.09}$   &    0.56 & $^{+0.10}_{-0.09}$  &     0.52  & $^{+0.05}_{-0.06}$   &    1.86  & $^{+0.37}_{-0.24}$ &  0.56  & $^{+0.08}_{-0.08}$   &   0.12   & $^{+0.08}_{-0.07}$  \\[3pt]
  K3   &    0.76  & $^{+0.28}_{-0.18}$  &    1.06  & $^{+0.11}_{-0.08}$   &    0.60 & $^{+0.10}_{-0.09}$  &     0.56  & $^{+0.05}_{-0.06}$   &    2.01  & $^{+0.34}_{-0.24}$ &  0.60  & $^{+0.07}_{-0.08}$   &   0.14   & $^{+0.07}_{-0.07}$  \\[3pt]
  K4   &    0.87  & $^{+0.25}_{-0.19}$  &    1.13  & $^{+0.11}_{-0.08}$   &    0.65 & $^{+0.10}_{-0.09}$  &     0.61  & $^{+0.06}_{-0.05}$   &    2.18  & $^{+0.35}_{-0.27}$ &  0.63  & $^{+0.07}_{-0.08}$   &   0.17   & $^{+0.07}_{-0.08}$  \\[3pt]
  K5   &    0.95  & $^{+0.26}_{-0.20}$  &    1.20  & $^{+0.10}_{-0.08}$   &    0.70 & $^{+0.09}_{-0.10}$  &     0.67  & $^{+0.06}_{-0.05}$   &    2.36  & $^{+0.34}_{-0.27}$ &  0.65  & $^{+0.07}_{-0.08}$   &   0.20   & $^{+0.06}_{-0.08}$  \\[3pt]
  K6   &    1.01  & $^{+0.28}_{-0.20}$  &    1.26  & $^{+0.08}_{-0.09}$   &    0.75 & $^{+0.09}_{-0.09}$  &     0.75  & $^{+0.06}_{-0.05}$   &    2.57  & $^{+0.33}_{-0.25}$ &  0.67  & $^{+0.07}_{-0.07}$   &   0.22   & $^{+0.06}_{-0.07}$  \\[3pt]
  K7   &    1.04  & $^{+0.25}_{-0.20}$  &    1.33  & $^{+0.07}_{-0.09}$   &    0.80 & $^{+0.08}_{-0.09}$  &     0.85  & $^{+0.07}_{-0.04}$   &    2.80  & $^{+0.33}_{-0.23}$ &  0.68  & $^{+0.07}_{-0.06}$   &   0.24   & $^{+0.05}_{-0.07}$  \\[3pt]
  M0   &     1.03 & $^{+0.25}_{-0.18}$   &    1.38 & $^{+0.06}_{-0.09}$    &    0.86& $^{+0.08}_{-0.08}$   &     0.96 & $^{+0.06}_{-0.04}$    &   3.07  & $^{+0.34}_{-0.21}$ &   0.69 & $^{+0.06}_{-0.06}$    &   0.25  & $^{+0.05}_{-0.07}$   \\[3pt]
  M1   &     1.01 & $^{+0.25}_{-0.14}$   &    1.42 & $^{+0.07}_{-0.08}$    &    0.92& $^{+0.08}_{-0.07}$   &     1.10 & $^{+0.06}_{-0.03}$    &   3.38  & $^{+0.33}_{-0.18}$ &   0.70 & $^{+0.05}_{-0.06}$    &   0.26  & $^{+0.05}_{-0.06}$   \\[3pt]
  M2   &     0.99 & $^{+0.25}_{-0.12}$   &    1.46 & $^{+0.08}_{-0.07}$    &    1.01& $^{+0.08}_{-0.06}$   &     1.26 & $^{+0.06}_{-0.03}$    &   3.75  & $^{+0.30}_{-0.16}$ &   0.69 & $^{+0.05}_{-0.06}$    &   0.27  & $^{+0.05}_{-0.06}$   \\[3pt]
  M3   &     0.99 & $^{+0.23}_{-0.10}$   &    1.51 & $^{+0.08}_{-0.05}$    &    1.12& $^{+0.09}_{-0.05}$   &     1.44 & $^{+0.06}_{-0.03}$    &   4.20  & $^{+0.30}_{-0.15}$ &   0.68 & $^{+0.05}_{-0.07}$    &   0.29  & $^{+0.05}_{-0.06}$   \\[3pt]
  M4   &     1.01 & $^{+0.21}_{-0.12}$   &    1.58 & $^{+0.08}_{-0.05}$    &    1.28& $^{+0.09}_{-0.05}$   &     1.63 & $^{+0.04}_{-0.04}$    &   4.75  & $^{+0.30}_{-0.13}$ &   0.66 & $^{+0.05}_{-0.06}$    &   0.30  & $^{+0.05}_{-0.06}$   \\[3pt]
  M5   &     1.05 & $^{+0.19}_{-0.16}$   &    1.68 & $^{+0.07}_{-0.06}$    &    1.48& $^{+0.09}_{-0.05}$   &     1.83 & $^{+0.03}_{-0.05}$    &   5.40  & $^{+0.28}_{-0.16}$ &   0.64 & $^{+0.06}_{-0.05}$    &   0.32  & $^{+0.05}_{-0.05}$   \\[3pt]
  M6   &     1.06 & $^{+0.18}_{-0.21}$   &    1.79 & $^{+0.06}_{-0.08}$    &    1.73& $^{+0.08}_{-0.07}$   &     2.02 & $^{+0.02}_{-0.06}$    &   6.13  & $^{+0.21}_{-0.20}$ &   0.62 & $^{+0.07}_{-0.04}$    &   0.34  & $^{+0.06}_{-0.05}$   \\[3pt]
  M7   &     1.04 & $^{+0.20}_{-0.25}$   &    1.90 & $^{+0.05}_{-0.10}$    &    2.01& $^{+0.08}_{-0.08}$   &     2.20 & $^{+0.03}_{-0.07}$    &   6.92  & $^{+0.23}_{-0.25}$ &   0.61 & $^{+0.08}_{-0.05}$    &   0.37  & $^{+0.07}_{-0.05}$   \\[3pt]
  M8   &     0.93 & $^{+0.40}_{-0.31}$   &    1.98 & $^{+0.07}_{-0.12}$    &    2.31& $^{+0.10}_{-0.11}$   &     2.36 & $^{+0.05}_{-0.08}$    &   7.72  & $^{+0.32}_{-0.29}$ &   0.63 & $^{+0.09}_{-0.07}$    &   0.42  & $^{+0.08}_{-0.06}$   \\[3pt]
  M9   &     0.69 & $^{+0.81}_{-0.37}$   &    2.02 & $^{+0.16}_{-0.14}$    &    2.62& $^{+0.19}_{-0.13}$   &     2.47 & $^{+0.10}_{-0.09}$    &   8.48  & $^{+0.65}_{-0.32}$ &   0.67 & $^{+0.12}_{-0.13}$    &   0.50  & $^{+0.10}_{-0.12}$   \\[3pt]
\hline

\end{tabular}
\tablefoot{For each color we report the best fit photospheric color at a given spectral type in the first column, and the +3$\sigma$ and $-3\sigma$ statistical uncertainty on this value in the second column. }

\end{center}
\end{table*}
%%%%%%%%%%%%%%%%%%%%%%%%%%%%%%%%%%%%%%%%%%%%%%%%%%%%%%

%%%%%%%%%%%%%%%%%%%%%%%%%%%%%%%%%%%%%
\begin{figure}[!t]
\centering
\includegraphics[width=0.5\textwidth]{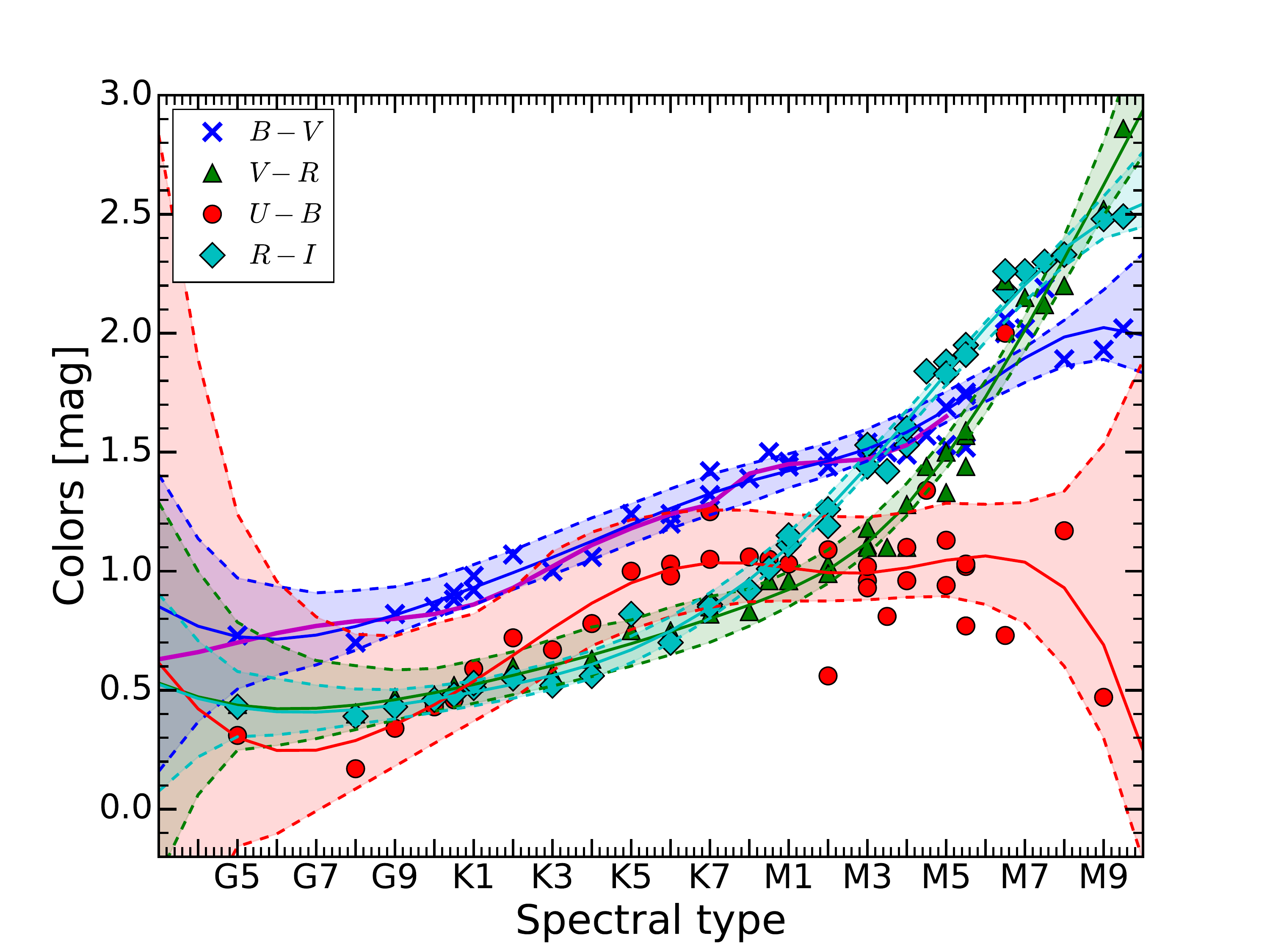}
\caption{Colors as a function of spectral type for the objects discussed here. The best fit for each color and the 3$\sigma$ confidence intervals are shown with a solid line and a shaded region. The solid purple line is the $B-V$ color for 5-30 Myr-old pre-main sequence stars by \citet{PM13}.
     \label{fig::phot_col_opt}}
\end{figure}
%%%%%%%%%%%%%%%%%%%%%%%%%%%%%%%%%%%%%%%%%%%%%%%%%%%%%%%%%%%%%%%%%%%%%%%%%%%%

%__________________________________________________________________

\section{Photospheric colors of young stars}\label{sect::photcol}
The absolutely flux-calibrated X-shooter spectra of our targets, with their  large, simultaneous  wavelength coverage and high signal-to-noise are perfectly suited to compute broad-band colors  for young stars and to be used to study variations of the extinction or the presence of spots on the surface of young objects \citep[e.g.,][]{Gully-Santiago17}. This kind of study relies on the knowledge of the intrinsic colors of the star, which are gravity dependent  \citep[e.g.,][]{Luhman03}. These spectra of non-accreting stars are thus ideal for calibrating the relation between spectral type and photospheric color for sub-giants.

\begin{figure}[!t]
\centering
\includegraphics[width=0.5\textwidth]{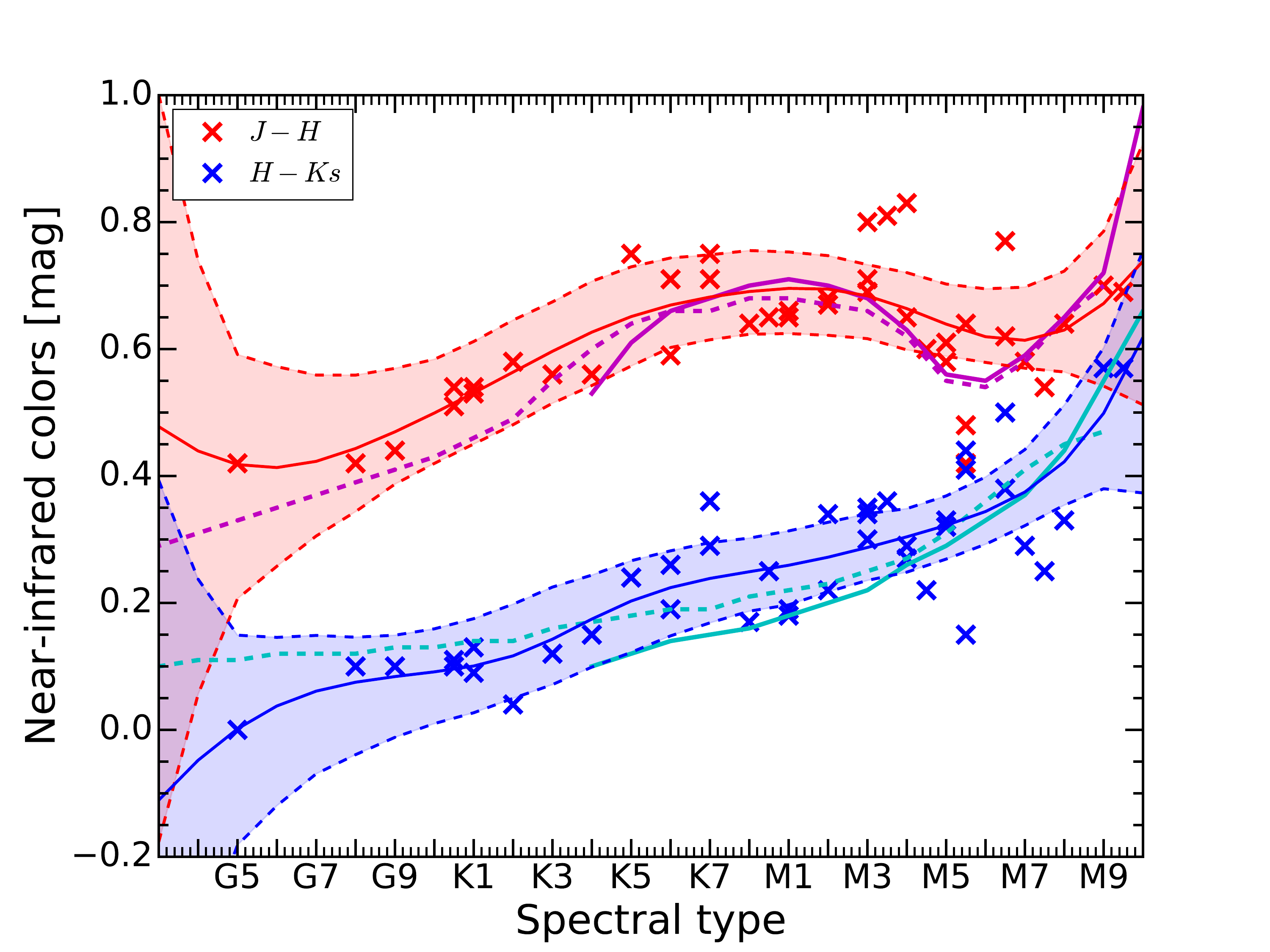}
\caption{Colors as a function of spectral type for the objects discussed here.  The best fit for each color and the 3$\sigma$ confidence intervals are shown with a solid line and a shaded region. The solid purple and cyan lines are the $J-H$ and $H-K$ photospheric colors for pre-main sequence stars by \citet{Luhman03}, while the dashed lines are the same colors for 5-30 Myr-old pre-main sequence stars by \citet{PM13}.
     \label{fig::phot_col_nir}}
\end{figure}

We perform synthetic photometry on the spectra in the following bands: Johnson $U$, $B$, and $V$, Cousins $R$ and $I$, and 2MASS $J$, $H$, $Ks$, using the throughputs provided with the TA-DA tool \citep{TADA}. We then plot the optical colors $B-V$, $V-R$, $U-B$, and $R-I$ for our targets as a function of their spectral type in Fig.~\ref{fig::phot_col_opt}, and similarly for the near-infrared $J-H$ and $H-K$ colors in Fig.~\ref{fig::phot_col_nir}. We show the $B-V$ color for pre-main sequence stars derived by \citet{PM13}, and the infrared colors by \citet{PM13} and \citet{Luhman03} in the plots. These photospheric colors are consistent within uncertainties with our estimate. We check that a value of $A_V$ larger than 0 mag, and up to 0.5 mag, would result in differences in the colors of up to 0.1 mag for the $U-B$ and $B-V$ colors, and less for the other colors. These differences are well within the uncertainties in the relation between spectral type and color we derive.
In general the colors increase for later spectral types, but the dependence on the spectral type is usually not a simple relation. For this reason, and in order not to generate bias due to prior assumption on the dependence of a color on the spectral type, we decided to use a non-parametric method to fit the data. In particular, we adopt a local second degree polynomial regression with a Gaussian kernel, using the Python module \textit{pyqt\_fit.npr\_methods.LocalPolynomialKernel}. We then bootstrap the result of this fit and derive the 3$\sigma$ confidence level for the fit, which is shown as a shaded region around the best fit on the plots. The most uncertain color is the $U-B$ color, with typical 3-$\sigma$ uncertainties of $\pm$0.3 mag. The largest uncertainty in the colors for later spectral types is due to the low S/N of the spectra, while for those at earlier spectral types the uncertainty is due to the low number of objects.

The relations between photospheric colors and spectral type for young stars, that is, sub-giants, calibrated on our spectra are reported in Table~\ref{tab::phot_col}. These are given as a function of spectral type since they are independent on the relation between spectral type and effective temperature. 

%%%%%%%%%%%%%%%%%%%%%%%%%%%%%%%%%%%%%
%%%%%%%%%%%%%%%%%%%%%%%%%%%%%%%%%%%%%%%%%%%%%%%%%%%%%%%%%%%%%%%%%%%%%%%%%%%%

%%%%%%%%%%%%%%%%%%%%%%%%%%%%%%%%%%%%%
\begin{figure}[!t]
\centering
\includegraphics[width=0.5\textwidth]{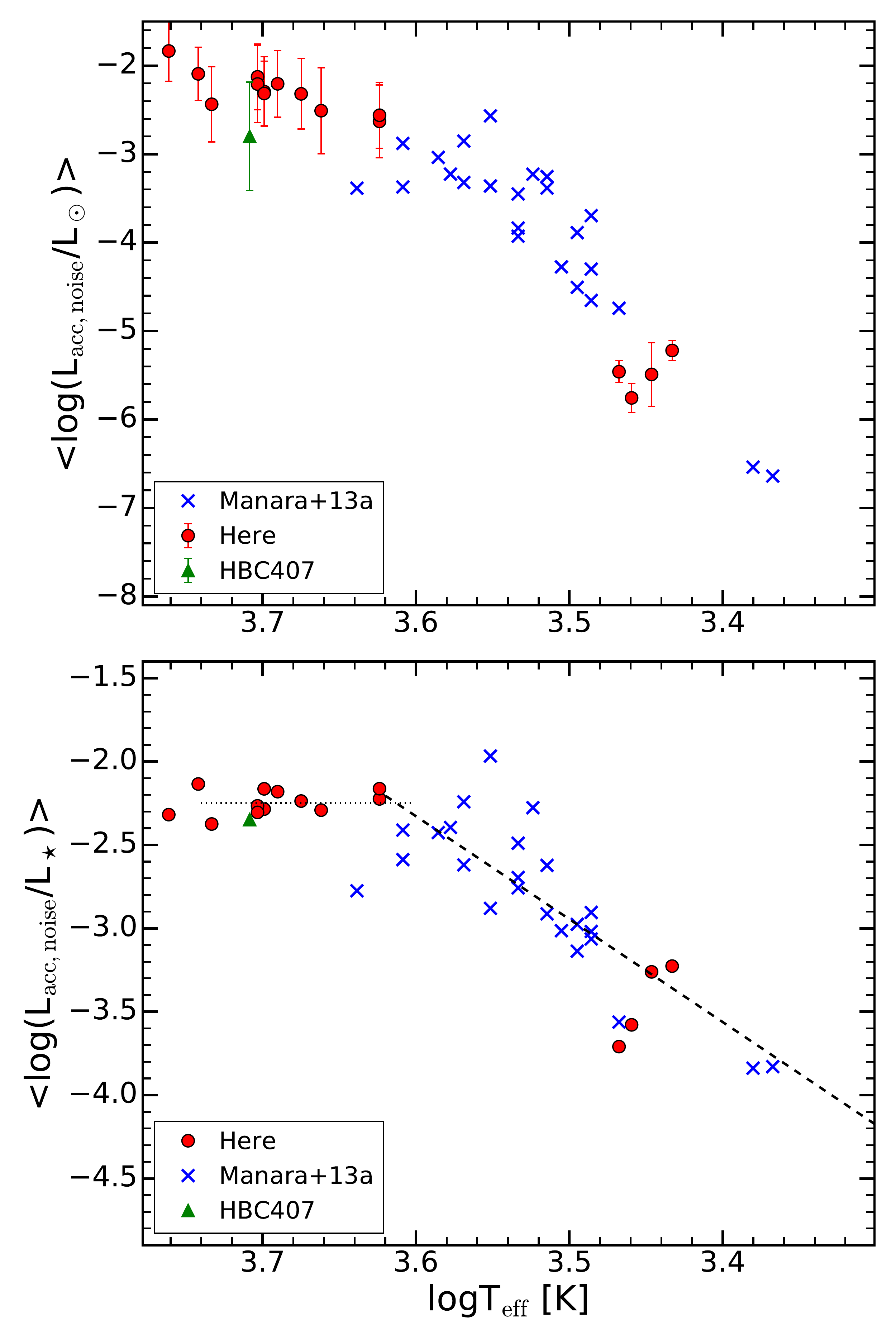}
\caption{Luminosity of chromospheric emission lines converted into \laccnoise \ as a function of the logarithm of the effective temperature obtained from the spectral type for all the ClassIII targets discussed here. In the bottom plot the values of \laccnoise \ are normalized by the stellar luminosity, and the lines represent the fit reported in Equation~\ref{eq::laccnoise}.
     \label{fig::laccnoise_vs_teff}}
\end{figure}
%%%%%%%%%%%%%%%%%%%%%%%%%%%%%%%%%%%%%%%%%%%%%%%%%%%%%%%%%%%%%%%%%%%%%%%%%%%%

%__________________________________________________________________

\section{Impact of chromospheric emission lines on accretion rate estimates}\label{sect::lines}

Young non-accreting stars typically show the well-known emission lines that characterize cromospheric activity; namely the hydrogen Balmer lines, most prominently the H$\alpha$ and H$\beta$ lines, the calcium lines, in particular the Ca II infrared triplet (CaIRT) lines at $\lambda\lambda\sim$  849.80 - 854.21 - 866.21 nm, the Ca K line at $\lambda\sim$ 393.37 nm and the Ca H line at $\lambda\sim$ 396.85 nm, and the helium lines, in particular the one at $\lambda\sim$ 587.6 nm. No emission lines due to chromospheric activity are typically present at infrared wavelengths. The chromospheric calcium lines usually appear as a narrow emission component inside a broad and pronounced photospheric absorption line, and similarly the hydrogen Balmer lines for objects hotter than $\sim$4000 K.  

The objects analyzed here span a large range of temperatures, and the presence of the aforementioned lines strongly depends on this parameter. The later-type objects with M spectral type in our sample show the H$\alpha$ and H$\beta$ lines in emission, with the only exception of LM601 where the H$\beta$ line is not detected possibly due to the low S/N of this spectrum at the wavelength of H$\beta$. None of these targets instead present emission in the calcium lines, not even when taking into account the photospheric absorption. The very low signal to noise ratio at the wavelengths of the H and K lines prevents detection of these lines for these cold targets, while the CaIRT lines are not detected even if the S/N is larger than 20 at the wavelengths of these transitions. We measure the flux of the H$\alpha$ and H$\beta$ lines in these targets by directly integrating the continuum-subtracted spectra, which leads to values consistent with those obtained by integrating the emission profiles in the spectrum obtained by the subtraction of the non-active template, since the underlying photospheric absorption is negligible in these cases.

On the other hand, the hotter objects with G and K spectral type in our sample always show prominent calcium and H$\alpha$ emission lines in reversal on the photospheric absorption lines. Thus, the subtraction of the photospheric template, taken as the best-fitting BT-Settl spectrum determined by ROTFIT, is mandatory to measure the flux of these lines. In three objects, namely RXJ1508.6-4423, RXJ1540.7-3756, and RXJ1543.1-3920, the H$\beta$ line is also detected after subtracting the photospheric absorption line, and we can measure their fluxes. In the following, we consider the flux from all the detected lines in the analysis, with the only exception being the Ca H line, as this is usually blended with that of H$\epsilon$ . Although the latter is expected to be faint, the decomposition of the profile is complicated given the projected rotational velocity of our targets and the resolution of our spectra, and we thus prefer to neglect this line in the analysis. The measured flux for these lines is reported in Table~\ref{tab::lines}.

Following \citetalias{Manara13a}, we apply the relations between line luminosity and accretion luminosity from \citet{Alcala14} to the luminosity of the lines produced in the chromospheres. As showed by \citet{Rigliaco12} and \citet{Alcala14}, the mean value of the accretion luminosity derived with these relations using multiple emission lines is consistent with the accretion luminosity obtained with the more direct method of directly measuring the excess emission in the Balmer continuum region. We thus want to obtain the equivalent accretion luminosity a typical chromosphere of a non-accreting young star would lead to by taking the mean value of as many emission lines as are present in the spectra. 
We call this equivalent accretion luminosity a ``noise'' on the accretion luminosity (\laccnoise).
We note that, after this conversion, the values of \laccnoise \ obtained using the H$\alpha$ line luminosity are systematically lower by $\sim$0.5 dex than those derived with the calcium lines, and $\sim$ 0.3 dex lower than those obtained using the H$\beta$ line. 
The discrepancy between the hydrogen Balmer and the calcium lines has already been reported by \citetalias{Manara13a}. The values of \laccnoise \ derived from the four calcium lines analyzed here are instead generally consistent with one another within $\sim$0.2 dex. The two exceptions to these general properties are RXJ1508.6-4423 and RXJ0445.8+1556. The former is one of the three hotter objects in our sample where the H$\beta$ line is detected. We see that the values of \laccnoise \ derived from the different lines for this object all agree with one another within $\sim$0.2 dex, while the values obtained using the Ca K line are higher than the mean value of \laccnoise \ by $\sim$0.7 dex. Similarly, the values of \laccnoise \ obtained for RXJ0445.8+1556 using the H$\alpha$ and CaIRT lines are discrepant by less than $\sim$0.3 dex, while the Ca K line flux leads to a value of \laccnoise \ higher by $\sim$ 0.7 dex. Since we are interested here in the typical impact of the typical values of the chromospheric emission on the measurements of accretion, and that in accreting objects this is better constrained when multiple lines are used, we decided to include all the detected lines in the analysis, and to use the mean value of \laccnoise \ of all these lines, which is reported in Table~\ref{tab::lines}. 

The dependence of \laccnoise \ on T$_{\rm eff}$, as well as the dependence of the stellar luminosity normalized \laccnoise \ on $T_{\rm eff}$ is shown in Fig.~\ref{fig::laccnoise_vs_teff}. The four newly added targets with late M spectral type follow the same linear trend in the log(\laccnoise/\lstar) versus log(T$_{\rm eff}$) relation as found by \citetalias{Manara13a}.  The relation  flattens at temperatures larger than $\sim$4000 K, and can be expressed as:  
\begin{multline}\label{eq::laccnoise}
\log(L_{\rm acc,noise}/L_\star) = -2.3 \pm 0.1 ~~~~~{\rm for~~~4000 < T_{eff} < 5800 K}\\
\log(L_{\rm acc,noise}/L_\star) = (6.2 \pm 0.5) \cdot \log T_{\rm eff} - (24.5 \pm 1.9) \\~~~~~{\rm for~~~ T_{eff} \le 4000 K},
\end{multline}

where the relation for low temperature is the one reported by \citetalias{Manara13a}. The values of $\log(L_{\rm acc,noise}/L_\star)$ have a small dependence on the assumed distance of the targets due to the fact that \laccnoise \ is not directly proportional to \lline. However, the value of \laccnoise \ reported for the objects with $T_{\rm eff} >$ 4000 K does not change even assuming the new distances from the Gaia TGAS catalog.

\begin{figure}[!t]
\centering
\includegraphics[width=0.5\textwidth]{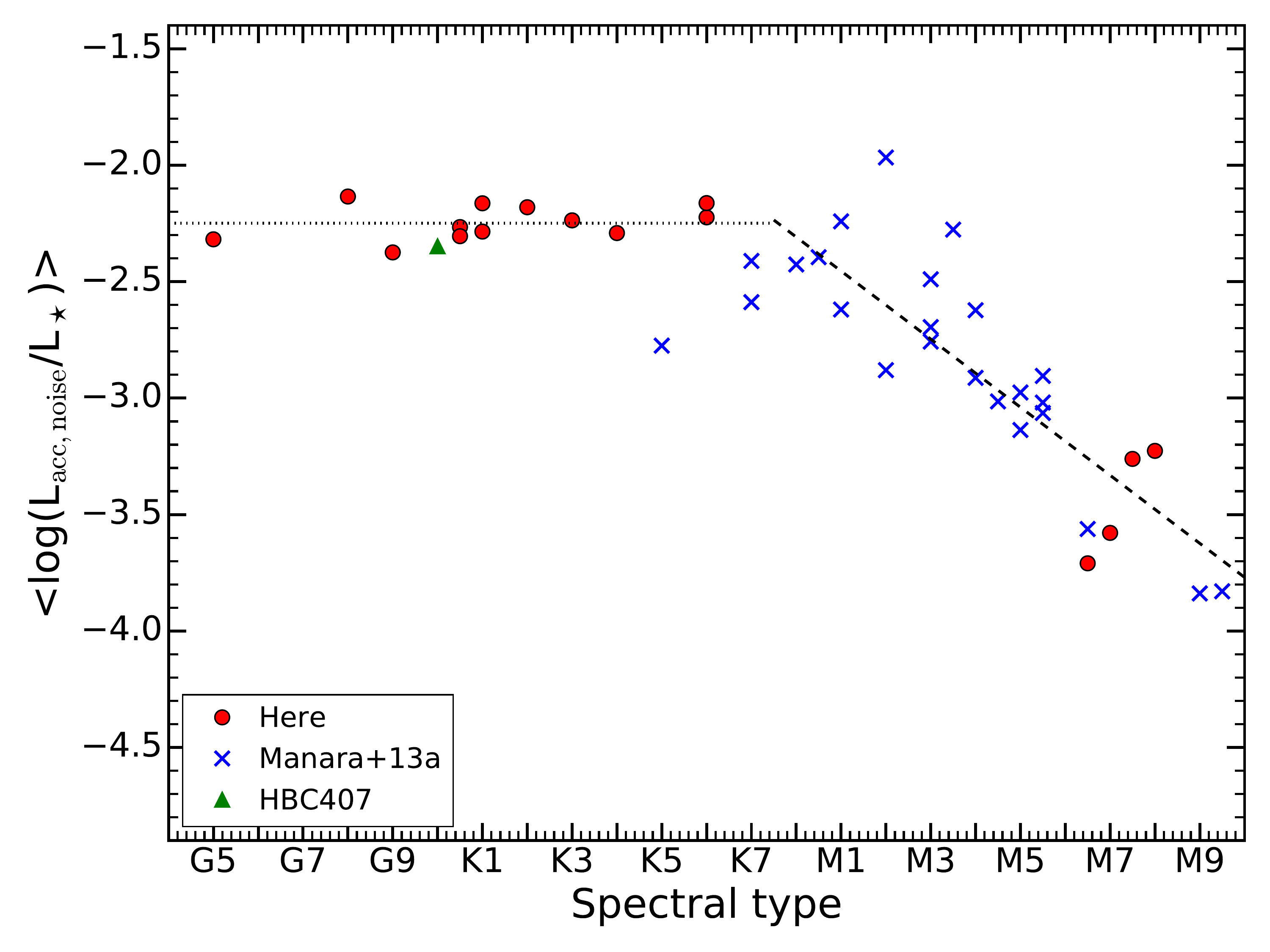}
\caption{Luminosity of chromospheric emission lines converted into \laccnoise \ and normalized by the stellar luminosity as a function of the spectral type for all the targets discussed here.
     \label{fig::laccnoise_vs_spt}}
\end{figure}

Since the relation between spectral type and $T_{\rm eff}$ for young stars is still debated in the literature (see Sect.~\ref{sect::rotfit}), it is useful to provide also a relation between log(\laccnoise/\lstar) and the spectral type, which is shown in Fig.~\ref{fig::laccnoise_vs_spt}. We perform the same analysis as for the dependence with $T_{\rm eff}$, and obtain the typical values of log(\laccnoise/\lstar) as a function of the spectral type of the targets, which are reported in Table~\ref{tab::laccnoise_spt}.

%%%%%%%%%%%%%%%%%%%%%%%%%%%%%%%%%%%%%
%%%%%%%%%%%%%%%%%%%%%%%%%%%%%%%%%%%%%%%%%%%%%%%%%%%%%%%%%%%%%%%%%%%%%%%%%%%%

%%%%%%%%%%%%%%%%%%%%%%%%%%%%%%%%%%%%%%%%%%%%%%%%%%%%%%
\begin{table}
\begin{center}
\footnotesize
\caption{\label{tab::laccnoise_spt} Typical noise on the measurements of accretion induced by chromospheric emission}
\begin{tabular}{l|c}  
\hline \hline

Spectral type & log(\laccnoise/\lstar)  \\

\hline

  G5   &   -2.25 $\pm$  0.07   \\
  G6   &   -2.25 $\pm$    0.07   \\
  G7   &   -2.25 $\pm$    0.07   \\
  G8   &   -2.25 $\pm$    0.07   \\
  G9   &   -2.25  $\pm$   0.07    \\
  K0   &   -2.25  $\pm$   0.07    \\
  K1   &   -2.25  $\pm$   0.07    \\
  K2   &   -2.25  $\pm$   0.07    \\
  K3   &   -2.25  $\pm$   0.07    \\
  K4   &   -2.25  $\pm$   0.07    \\
  K5   &   -2.25  $\pm$   0.07    \\
  K6   &   -2.25  $\pm$   0.07    \\
  K7   &   -2.25  $\pm$   0.07    \\
  M0   &   -2.31   $\pm$  0.07     \\
  M1   &   -2.45   $\pm$  0.08     \\
  M2   &   -2.60   $\pm$  0.09     \\
  M3   &   -2.75   $\pm$  0.11     \\
  M4   &   -2.89   $\pm$  0.12     \\
  M5   &   -3.04   $\pm$  0.13     \\
  M6   &   -3.19   $\pm$  0.15     \\
  M7   &   -3.33   $\pm$  0.16     \\
  M8   &   -3.48   $\pm$  0.17     \\
  M9   &   -3.62   $\pm$  0.19     \\
\hline

\end{tabular}
\tablefoot{The uncertainty reported for $\log L_{\rm acc, noise}$ is the 1$\sigma$ uncertainty.  }

\end{center}
\end{table}
%%%%%%%%%%%%%%%%%%%%%%%%%%%%%%%%%%%%%%%%%%%%%%%%%%%%%%

The flattening of the relation for objects of spectral type earlier than about K7 was already noticed by \citetalias{Manara13a}, whose sample, however, included only a couple of stars earlier than K7. This is possibly related to the saturation of
the chromospheric activity in hotter stars \citep[see e.g.,][]{Stelzer13,Frasca15}. However, we want to stress that  \laccnoise, computed following the same procedure used to measure \lacc \ in accreting stars, that is,  averaging the values of \lacc \ derived from each detected emission line, is not an appropriate diagnostic of chromospheric activity, but is only a way to estimate the effect of chromospheric emission on the measurements of accretion rates. The analysis of the chromospheric properties of these objects is not among the aims of this work.

%__________________________________________________________________

\subsection{Implications for accretion rate estimates}

\begin{figure}[!t]
\centering
\includegraphics[width=0.5\textwidth]{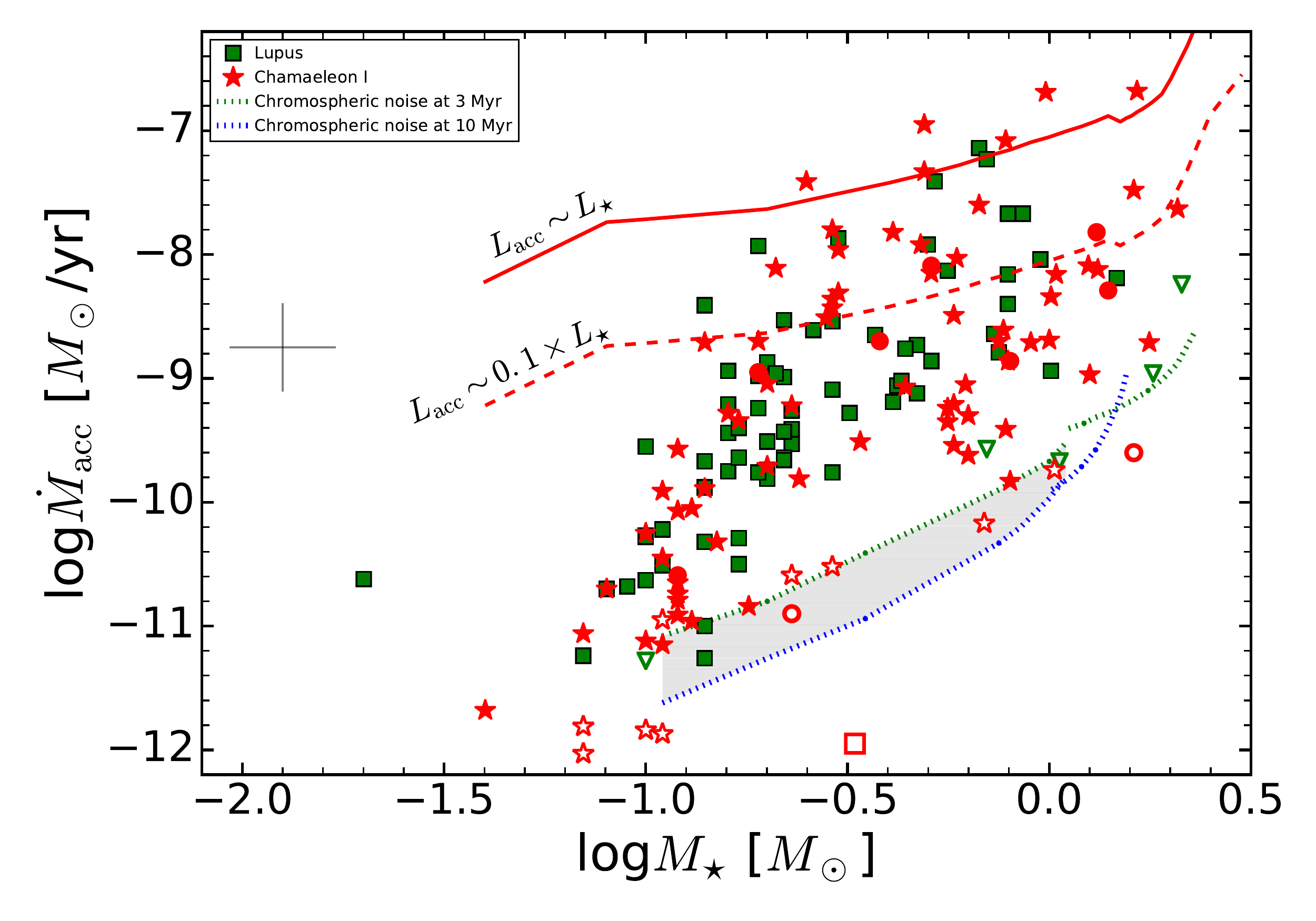}
\caption{Compilation of mass accretion rates vs. stellar masses measured from VLT/X-Shooter spectra by \citet[][green squares, Lupus sample]{Alcala14,Alcala17} and by \citet[][red stars, Chamaeleon~I sample]{Manara16a,Manara17}. 
The red solid and dashed lines show the values of mass accretion rates corresponding to objects whose accretion luminosity is respectively equal to the stellar luminosity or to 10\% of it. The limits derived here due to chromospheric emission are shown with green and blue dotted lines for $\sim$3 Myr and $\sim$10 Myr old objects, respectively.
     \label{fig::macc_mstar}}
\end{figure}

The chromospheric emission of young stars derived here impacts the measurements of \lacc \ in accreting young stars and, in turn, of \macc. It is particularly instructive to find the values of \macc \ at a given \mstar \ corresponding to the typical chromospheric emission. We thus use the value of \laccnoise \ for objects with \mstar $>$ 1 \msun \ from Eq.~\ref{eq::laccnoise} and convert this into \macc \ adopting the stellar luminosity, mass, and radius from one isochrone of the evolutionary models by \citet{Siess00}, similarly as done by \citetalias{Manara13a} for lower-mass stars. This model is chosen as it extends to higher stellar masses than the more recent models by \citet{Baraffe15}, and it is thus more appropriate for this analysis. 
The resulting values of \macc, which are the limit of the detectable accretion rate in chromospherically active young stars, are shown in Fig.~\ref{fig::macc_mstar} for the 3 Myr and 10 Myr isochrones by \citet{Siess00}, and together with the same limits derived by \citetalias{Manara13a} for \mstar$<$1 \msun. In this plot we also show the measurements of \macc \ for the close to complete samples of Class II  stars (i.e., with evidence of disks) in the Lupus \citep{Alcala14,Alcala17} and Chamaeleon~I \citep{Manara16a,Manara17} star-forming regions. The majority of targets show measured accretion rates well above the chromospheric limits. However, both samples include some Class II objects whose measured accretion rate is compatible or smaller than the expected chromospheric emission, and these are shown with empty symbols. The measured emission for these targets is thus probably dominated by chromospheric activity, and it is not possible to determine whether they are accreting or not. For this reason these targets have been defined as ``dubious accretors'' by \citet{Alcala17} and \citet{Manara17}. 

%%%%%%%%%%%%%%%%%%%%%%%%%%%%%%%%%%%%%
%%%%%%%%%%%%%%%%%%%%%%%%%%%%%%%%%%%%%%%%%%%%%%%%%%%%%%%%%%%%%%%%%%%%%%%%%%%%

%%%%%%%%%%%%%%%%%%%%%%%%%%%%%%%%%%%%%%%%%%%%%%%%%%%%%%
\begin{table*}
\begin{center}
\footnotesize
\caption{\label{tab::lines} Flux of emission lines and chromospheric noise}
\begin{tabular}{l|cc|cccc|c}  
\hline \hline
 Object & $F_{\rm H\alpha}$ & $F_{\rm H\beta}$ & $F_{\rm CaK 393.4 nm}$ &  $F_{\rm Ca 849.8 nm}$ & $F_{\rm Ca 854.2 nm}$ & $F_{\rm Ca 866.2 nm}$ & $\log L_{\rm acc, noise}$ \\

\hline

    RXJ0445.8+1556 &    (9.0$\pm$1.3)$\cdot 10^{-13}$ & \nodata & (6.9$\pm$1.3)$\cdot 10^{-13}$ & (7.6$\pm$1.7)$\cdot 10^{-14}$  &  (6.6$\pm$1.4)$\cdot 10^{-14}$  &  (8.7$\pm$1.7)$\cdot 10^{-14}$ & -1.8$\pm$0.3 \\
    RXJ1526.0-4501 &    (8.5$\pm$1.7)$\cdot 10^{-14}$ & \nodata &(9.1$\pm$3.1)$\cdot 10^{-14}$ & (2.1$\pm$0.3)$\cdot 10^{-14}$  &  (3.2$\pm$0.4)$\cdot 10^{-14}$  &  (2.7$\pm$0.3)$\cdot 10^{-14}$ & -2.4$\pm$0.4 \\
 PZ99J160550.5-253313 & (1.3$\pm$0.2)$\cdot 10^{-13}$ & \nodata & (8.6$\pm$3.6)$\cdot 10^{-14}$  & (4.2$\pm$0.5)$\cdot 10^{-14}$  &  (5.4$\pm$0.6)$\cdot 10^{-14}$  &  (4.4$\pm$0.5)$\cdot 10^{-14}$ & -2.3$\pm$0.4 \\
    RXJ1508.6-4423 &    (6.6$\pm$0.7)$\cdot 10^{-13}$ & (1.6$\pm$0.3)$\cdot 10^{-13}$ & (3.2$\pm$0.6)$\cdot 10^{-13}$ & (2.8$\pm$0.6)$\cdot 10^{-14}$  &  (3.1$\pm$0.5)$\cdot 10^{-14}$  &  (3.3$\pm$0.6)$\cdot 10^{-14}$ & -2.1$\pm$0.3 \\
 PZ99J160843.4-260216 & (2.0$\pm$0.3)$\cdot 10^{-13}$ & \nodata & (1.6$\pm$0.4)$\cdot 10^{-13}$  & (5.0$\pm$0.6)$\cdot 10^{-14}$  &  (7.1$\pm$0.8)$\cdot 10^{-14}$  &  (6.2$\pm$0.7)$\cdot 10^{-14}$ & -2.2$\pm$0.4 \\
    RXJ1515.8-3331 &    (1.2$\pm$0.2)$\cdot 10^{-13}$ & \nodata & (1.3$\pm$0.4)$\cdot 10^{-13}$  & (4.2$\pm$0.5)$\cdot 10^{-14}$  &  (6.1$\pm$0.7)$\cdot 10^{-14}$  &  (5.4$\pm$0.6)$\cdot 10^{-14}$ & -2.2$\pm$0.4 \\
    RXJ1547.7-4018 &    (1.1$\pm$0.2)$\cdot 10^{-13}$ & \nodata & (8.9$\pm$3.3)$\cdot 10^{-14}$ & (3.1$\pm$0.4)$\cdot 10^{-14}$  &  (4.5$\pm$0.5)$\cdot 10^{-14}$  &  (3.9$\pm$0.4)$\cdot 10^{-14}$ & -2.3$\pm$0.4 \\
    RXJ0438.6+1546 &    (1.7$\pm$0.2)$\cdot 10^{-13}$ & \nodata & (1.0$\pm$0.2)$\cdot 10^{-13}$ & (5.4$\pm$0.6)$\cdot 10^{-14}$  &  (7.3$\pm$0.8)$\cdot 10^{-14}$  &  (6.0$\pm$0.6)$\cdot 10^{-14}$ & -2.2$\pm$0.4 \\
    RXJ0457.5+2014 &    (1.5$\pm$0.2)$\cdot 10^{-13}$ & \nodata & (9.4$\pm$2.5)$\cdot 10^{-14}$  & (3.7$\pm$0.5)$\cdot 10^{-14}$  &  (5.2$\pm$0.6)$\cdot 10^{-14}$  &  (4.4$\pm$0.6)$\cdot 10^{-14}$ & -2.3$\pm$0.4 \\
    RXJ1538.6-3916 &    (5.2$\pm$0.9)$\cdot 10^{-14}$ & \nodata & (5.5$\pm$1.0)$\cdot 10^{-14}$ & (2.3$\pm$0.3)$\cdot 10^{-14}$  &  (3.2$\pm$0.4)$\cdot 10^{-14}$  &  (3.0$\pm$0.3)$\cdot 10^{-14}$ & -2.5$\pm$0.5 \\
    RXJ1540.7-3756 &    (6.4$\pm$0.9)$\cdot 10^{-14}$ & (2.3$\pm$1.0)$\cdot 10^{-14}$ & (3.7$\pm$0.5)$\cdot 10^{-14}$  & (2.2$\pm$0.2)$\cdot 10^{-14}$  &  (3.0$\pm$0.3)$\cdot 10^{-14}$  &  (2.4$\pm$0.3)$\cdot 10^{-14}$ & -2.6$\pm$0.4 \\
    RXJ1543.1-3920 &    (7.5$\pm$1.0)$\cdot 10^{-14}$ & (3.5$\pm$1.8)$\cdot 10^{-14}$ & (4.0$\pm$0.4)$\cdot 10^{-14}$ & (2.3$\pm$0.2)$\cdot 10^{-14}$  &  (3.1$\pm$0.3)$\cdot 10^{-14}$  & (2.7$\pm$0.3)$\cdot 10^{-14}$ & -2.6$\pm$0.4 \\
\hline
                 LM717 &  (9.9$\pm$0.5)$\cdot 10^{-16}$  &  (9.7$\pm$1.0)$\cdot 10^{-17}$  & \nodata  & \nodata   &   \nodata    &   \nodata & -5.5$\pm$0.1 \\
                 LM601 &  (3.4$\pm$0.3)$\cdot 10^{-16}$  &  $<$1.352$\cdot 10^{-16}$ & \nodata   & \nodata   &  \nodata   &    \nodata  & -5.5$\pm$0.4 \\
     J11195652-7504529 &  (5.8$\pm$0.4)$\cdot 10^{-16}$ & (4.8$\pm$1.6)$\cdot 10^{-17}$ & \nodata  & \nodata  &    \nodata    &   \nodata  & -5.8$\pm$0.2 \\
             CHSM17173 &  (1.59$\pm$0.06)$\cdot 10^{-15}$ & (1.6$\pm$0.2)$\cdot 10^{-16}$ & \nodata  & \nodata    &   \nodata    &   \nodata  & -5.2$\pm$0.1 \\
\hline

 HBC407 &   (3.1$\pm$0.7)$\cdot 10^{-14}$ & \nodata & \nodata & (1.1$\pm$0.1)$\cdot 10^{-14}$   & (1.6$\pm$0.2)$\cdot 10^{-14}$ &   (1.5$\pm$0.2)$\cdot 10^{-14}$  & -2.8$\pm$0.6 \\

\hline

\end{tabular}
\tablefoot{Fluxes are in  [erg $\cdot$ s$^{-1} \cdot$ cm$^{-2}$]. The uncertainty reported for $\log L_{\rm acc, noise}$ is the standard deviation of the values of log$L_{\rm acc,noise}$ measured with the different lines. For LM601 we report the uncertainty on the values of \laccnoise \ measured from the H$\alpha$ line.  }

\end{center}
\end{table*}
%%%%%%%%%%%%%%%%%%%%%%%%%%%%%%%%%%%%%%%%%%%%%%%%%%%%%%

%
%
%
%%__________________________________________________________________
%
\section{Summary and conclusions}

We present the analysis of new VLT/X-Shooter spectra of photospheric templates of young stars that complements the previous analysis by \citet{Manara13a} by including objects with spectral types not covered in the previous work. All the spectra are available in reduced form on the CDS website, and they represent a complete set of photospheric templates of young stars with spectral type from G5 to M9.5 and with typically two or more spectra per spectral subclass. 
The spectra have all been analyzed homogeneously, with spectral types obtained by means of spectral indices and by comparing the relative strength of photospheric absorption features, as well as with an analysis with the ROTFIT tool to derive the photospheric properties ($T_{\rm eff}$, log$g$, $v$sin$i$, RV). We have shown that the two methods produce results in agreement with each other, and that the photospheric properties are compatible with the young ages of these objects. We see the presence of an apparent mass-related isochronal age dependence (younger age for lower-mass stars) when comparing the positions of our targets on the HRD with evolutionary models, as already noted in the literature.

Thanks to the broad wavelength coverage of our spectra and to their absolute flux calibration, we have derived photospheric colors in multiple bands ($U, B, V, R, I, J, H, K_s$) as a function of the spectral type. These colors are the first multi-band set of colors computed for a set of pre-main sequence stars, that is, sub-giants, with an ample coverage of spectral types and can be used as standard colors for this kind of object.

We have measured the flux of several emission lines due to chromospheric activity by properly subtracting the photospheric absorption lines. We have converted the flux of these lines into \laccnoise, which is the equivalent value of accretion luminosity one would obtain if this emission were due to accretion processes in young stars with disks. This value, which represents the typical lower limit on the measurable accretion rates in young stars, is independent to the stellar temperature for $T_{\rm eff}>4000$ K when normalized for the stellar luminosity. This is possibly related to the saturation of chromospheric emission in these young objects. On the other hand, our data confirm the results of \citet{Manara13a} that the values of \laccnoise \ decrease with decreasing stellar temperature for $T_{\rm eff}<4000$ K. We also provide the typical values of \laccnoise \ for spectral types ranging from G5 to M9.
When converted into mass accretion rates, this limit indicates that objects at ages of $\sim$ 1-5 Myr with measured accretion rate $\lesssim 3\cdot 10^{-10} M_\odot$/yr and stellar masses $\sim1.5 M_\odot$ are probably dominated by chromospheric emission and should be considered with caution.

\begin{acknowledgements}
We are grateful to the anonymous referee for the detailed and constructive report which allowed us to  improve the quality of the paper.
      CFM gratefully acknowledges an ESA Research Fellowship. A.N. acknowledges support from Science Foundation Ireland (Grant 13/ERC/12907). Financial support from INAF under the program  PRIN2013 ``Disk jets and the dawn of planets'' is also acknowledged. We also thank the ESO and the Paranal staff for their support during phase-2 proposal preparation and for the observations. This research made use of Astropy, SIMBAD, and matplotlib. We are thankful to F. Allard and T. Prusti for inspiring discussions.
\end{acknowledgements}

% WARNING
%-------------------------------------------------------------------
% Please note that we have included the references to the file aa.dem in
% order to compile it, but we ask you to:
%
% - use BibTeX with the regular commands:
%   \bibliographystyle{aa} % style aa.bst
%   \bibliography{Yourfile} % your references Yourfile.bib
%
% - join the .bib files when you upload your source files
%-------------------------------------------------------------------

\clearpage

\appendix

\section{Additional data}\label{app::lit}

Additional information from the literature on the targets discussed here is reported in Tables~\ref{tab::lit}, and archival photometric data in Tables~\ref{tab::phot}. Finally, the observing log is reported in Table~\ref{tab::log}.

%%%%%%%%%%%%%%%%%%%%%%%%%%%%%%%%%%%%%%%%%%%%%%%%%%%%%%
% \setlength{\tabcolsep}{3pt}
% version of 2015-01-05
\begin{table*}
\begin{center}
\footnotesize
\caption{\label{tab::lit} Data available in the literature for the Class~III YSOs included in this work }
\begin{tabular}{l|cc|c|cc|c|c}  
\hline \hline
 Object/other name &  RA(2000)  & DEC(2000) &  Region & SpT & A$_V$ &  Notes & References \\ 
      &  h \, :m \, :s & $^\circ$ \, ' \, ''   &  & \hbox{} & [mag] & \hbox{} & \hbox{} \\         
\hline

RXJ0445.8+1556  &               04:45:51.29 &+15:55:49.69       &Taurus & G5      &       0.2     &       not SB          & 1,6 \\
RXJ1526.0-4501  &               15:25:59.65 &-45:01:15.72       &Lupus & G5/G7   &       0.4     &not SB & 1\\
PZ99J160550.5-253313    &       16:05:50.64& -25:33:13.60 &Upper Sco & G7/K1 & 0.7   &\nodata        &1 \\
RXJ1508.6-4423  &               15:08:37.75& -44:23:16.95&Lupus &       G8      &       0.0     & not SB & 1\\
PZ99J160843.4-260216            &16:08:43.41 &-26:02:16.84      &Upper Sco &G9/G7  &       0.7     & not SB        &1\\
RXJ1515.8-3331          &       15:15:45.36 &-33:31:59.78&      Lupus & K0      &       0.0 & \nodata & 1\\
RXJ1547.7-4018  &               15:47:41.76 &-40:18:26.80&      Lupus & K1      &       0.1     &\nodata & 1\\
RXJ0438.6+1546          &       04:38:39.07& +15:46:13.61&      Taurus & K2      &       0.2     &not binary     &1,6,7\\
RXJ0457.5+2014          &       04:57:30.66 &+20:14:29.42& Taurus &     K3      &       0.4     & 6.9\arcsec~binary and SB1 &1,6\\
RXJ1538.6-3916          &       15:38:38.36 &-39:16:54.08&Lupus &       K4      &       0.4     &not SB      &1\\RXJ1540.7-3756              &       15:40:41.17& -37:56:18.54&      Lupus & K6    &       0.1     &       not SB          &1\\
RXJ1543.1-3920          &       15:43:06.25 & -39:20:19.5&Lupus &       K6      &       0.1& \nodata &1\\
\hline
LM717           &               11:08:02.34 &-76:40:34.3                & Chamaeleon I & M6       &       0.4     &\nodata &2,3\\
LM601                   &       11:12:30.99& -76:53:34.2                &Chamaeleon I &M7           &0.0    &\nodata &2,3\\
J11195652-7504529       &       11:19:56.52 &-75:04:52.9        &       Chamaeleon I &M7.25        &       0.0     &\nodata &2,3\\
CHSM17173               &       11:10:22.26 &-76:25:13.8                &Chamaeleon I &M8   &       0.0     &\nodata & 2,3\\

\hline

\end{tabular}
\tablefoot{ Spectral types, extinction, disk classification, accretion indication, and binarity are adopted from the following studies: 1.~\citet{Wahhaj10}, 2.~\citet{Luhman07}; 3.~\citet{Luhman08}; 4.~\citet{Luhman04}; 5.~\citet{Manoj11}; 6.~\citet{Nguyen12}; 7.~\citet{Daemgen15}. SB stands for spectroscopic binary. }

\end{center}
\end{table*}
%%%%%%%%%%%%%%%%%%%%%%%%%%%%%%%%%%%%%%%%%%%%%%%%%%%%%%

%%%%%%%%%%%%%%%%%%%%%%%%%%%%%%%%%%%%%%%%%%%%%%%%%%%%%
% \setlength{\tabcolsep}{3pt}
\begin{table*}
\begin{center}
\footnotesize
\caption{\label{tab::phot} Photometry available in the literature for the Class~III YSOs included in this work }
\begin{tabular}{l|cccccccc}  
\hline \hline
 Object &$U$&$B$&$      V       $&$R$&$ I$&$    J$&$    H$&$    K$ \\ 
       & [mag]& [mag]& [mag]& [mag]& [mag]& [mag]& [mag]& [mag] \\         
\hline

RXJ0445.8+1556  &       \nodata &10.12& 9.36    &\nodata        &8.43&  7.8     &7.46&  7.34 \\
RXJ1526.0-4501          &       11.97&  11.67&  10.97&  10.42&  10.04&  9.4&    8.98    &8.90   \\
PZ99J160550.5-253313    &\nodata        &11.87& 10.89&  \nodata&        9.80    &9.1    &8.59   &8.46\\
RXJ1508.6-4423          &       11.46   &11.54& 10.65   &10.22& 9.80    &9.36&  8.92    &8.81   \\
PZ99J160843.4-260216            &       \nodata &11.16& 10.38&  \nodata&        9.27    &8.55&  8.05&   7.91    \\
RXJ1515.8-3331          &       12.00&  11.61&  10.67   &10.17& 9.57    &8.98&  8.46&   8.38    \\
RXJ1547.7-4018          &       12.58   &12.33  &11.16& 10.75   &10.05& 9.29&   8.81    &8.66\\
RXJ0438.6+1546          &       \nodata &11.75& 10.86&  \nodata &9.62   &8.90&  8.36    &8.24\\
RXJ0457.5+2014          &       \nodata &11.99& 11.06&  \nodata&        9.94    &9.28   &8.82&  8.69\\
RXJ1538.6-3916          &\nodata        &\nodata        &11.47& \nodata&        \nodata&        9.59    &9.01&  8.85    \\
RXJ1540.7-3756          &       \nodata &13.10& 12.35&  11.70&  \nodata &9.93   &9.32   &9.19   \\
RXJ1543.1-3920          &                       \nodata &13.70  &12.25  &11.50  &\nodata&       9.85&   9.22&   9.10    \\
\hline
LM717                   &       \nodata &\nodata        &\nodata        &18.21& 15.46   &12.95& 12.31   &11.95\\
LM601                   &       \nodata &\nodata        &\nodata        &\nodata        &\nodata        &14.07& 13.51&  13.05\\
J11195652-7504529&\nodata&      \nodata &\nodata        &\nodata        &16.68  &14.05& 13.33   &12.98\\
CHSM17173                       &       \nodata&        18.7&   \nodata &16.80  &16.58  &13.53  &12.90  &12.47  \\

\hline

\end{tabular}

\end{center}
\end{table*}
%%%%%%%%%%%%%%%%%%%%%%%%%%%%%%%%%%%%%%%%%%%%%%%%%%%%%

%%%%%%%%%%%%%%%%%%%%%%%%%%%%%%%%%%%%%%%%%%%%%%%%%%%%%
\begin{table*}  
\begin{center}  
\footnotesize  
\caption{\label{tab::log} Night log of the observations }  
\begin{tabular}{l|c|ccc| cc | c }    
\hline \hline  
 Name  &  Date of observation [UT] & \multicolumn{3}{c}{Slit Width [\arcsec x11\arcsec]} | &  \multicolumn{2}{c}{Exp. Time [Nexp x (s)]} |& Exp. Time [Nexp x NDIT x (s)] \\    
   &   &  UVB & VIS & NIR & UVB & VIS & NIR  \\    
\hline  

RXJ0445.8+1556                          & 2015-10-17T05:04:37.658 &     0.5 &       0.4 &   0.4 &   2x180 &         2x90 &  2x3x50 \\       
RXJ1526.0-4501                          & 2016-02-17T07:31:01.419 &     0.5 &       0.4 &   0.4 &   2x180 &         2x90 &  2x3x50 \\       
PZ99J160550.5-253313                    & 2016-01-28T09:09:47.254 &     0.5 &       0.4 &   0.4 &   2x190 &         2x100 &         2x3x60 \\
RXJ1508.6-4423                          & 2016-02-17T08:02:13.669 &     0.5 &       0.4 &   0.4 &   2x180 &         2x90 &  2x3x50 \\       
PZ99J160843.4-260216                    & 2016-03-21T04:36:57.171 &     0.5 &       0.4 &   0.4 &   2x180 &         2x90  &         2x3x50 \\       
RXJ1515.8-3331                          & 2016-02-17T08:53:32.989 &     0.5 &       0.4 &   0.4 &   2x180 &         2x90 &  2x3x50 \\       
RXJ1547.7-4018                          & 2016-02-11T08:57:22.320 &     0.5 &       0.4 &   0.4 &   2x220 &         2x130 &         2x3x60 \\       
RXJ0438.6+1546                          & 2015-10-17T05:46:35.986 &     0.5 &       0.4 &   0.4 &   2x180 &         2x90 &  2x3x50 \\       
RXJ0457.5+2014                          & 2015-10-28T07:43:56.029 &     0.5 &       0.4 &   0.4 &   2x180 &         2x90 &  2x3x50 \\               
RXJ1538.6-3916                          & 2016-02-16T08:36:51.577 &     0.5 &       0.4 &   0.4 & 4x200 & 4x100 & 4x3x50 \\         
RXJ1540.7-3756                          & 2016-03-03T06:06:55.329 &     0.5 &       0.4 &   0.4 & 4x220 & 4x130 & 4x3x60 \\         
RXJ1543.1-3920                          & 2016-02-17T08:20:47.360 &     0.5 &       0.4 &   0.4 & 4x250 & 4x150 & 4x3x50 \\

\hline
LM717                                   & 2016-01-28T06:10:11.765 &     1.3 &       0.9 &   0.9 & 4x900 & 4x810 & 4x3x300 \\        
LM601                                   & 2016-01-26T05:14:58.117 &     1.3 &       0.9 &   0.9 & 4x900 & 4x810 & 4x3x300 \\                
J11195652-7504529                       & 2016-01-27T04:55:37.169 &     1.3 &       0.9 &   0.9 & 4x900 & 4x810 & 4x3x300 \\                
CHSM17173                               & 2016-01-28T03:28:16.115 &     1.3 &       0.9 &   0.9 & 4x900 & 4x810 & 4x3x300 \\

\hline 
\end{tabular} 
\end{center} 
\end{table*}  
%%%%%%%%%%%%%%%%%%%%%%%%%%%%%%%%%%%%%%%%%%%%%%%%%%%%%

\section{Values of the $F_{\rm red}$ ratio as a function of extinction and spectral type}

%%%%%%%%%%%%%%%%%%%%%%%%%%%%%%%%%%%%%%%%%%%%%%%%%%%%%%%%%%%%%%%%%%%%%%%%%%%%
\begin{table*}  
\begin{center}  
\footnotesize  
\caption{\label{tab::flred} Values of the ratio $F_{\rm 833 nm}/F_{\rm 634.8 nm}$ }  
\begin{tabular}{l|ccccccccc }    
\hline \hline  

 Sp. Type &   \multicolumn{9}{c}{$F_{\rm red} = F_{\rm 833 nm}/F_{\rm 634.8 nm}$}\\
\hbox{} & $A_V$ = 0 mag & $A_V$ = 0.5 mag & $A_V$ = 1 mag  & $A_V$ = 2 mag  &  $A_V$ = 3 mag  &  $A_V$ = 4 mag  &  $A_V$ = 5 mag  & $A_V$ = 6 mag  & $A_V$ = 10 mag \\

\hline
  G5    & 0.682      &     0.777     &      0.886    &       1.151     &      1.496    &       1.940    &       2.515      &     3.262    &       9.256 \\         
  G6    & 0.688      &     0.784     &      0.893    &       1.160     &      1.508    &       1.956    &       2.536      &     3.290    &       9.331 \\         
  G7    & 0.694      &     0.791     &      0.901    &       1.170     &      1.520    &       1.972    &       2.557      &     3.317    &       9.405 \\         
  G8    & 0.700      &     0.798     &      0.909    &       1.179     &      1.532    &       1.987    &       2.579      &     3.345    &       9.480 \\         
  K0    & 0.712      &     0.812     &      0.925    &       1.199     &      1.556    &       2.018    &       2.621      &     3.400    &       9.629 \\         
  K1    & 0.718      &     0.818     &      0.932    &       1.208     &      1.568    &       2.034    &       2.642      &     3.428    &       9.703 \\         
  K2    & 0.725      &     0.825     &      0.940    &       1.218     &      1.581    &       2.049    &       2.663      &     3.455    &       9.777 \\         
  K3    & 0.778      &     0.886     &      1.009    &       1.308     &      1.694    &       2.197    &       2.850      &     3.697    &       10.481 \\        
  K4    & 0.842      &     0.959     &      1.091    &       1.415     &      1.833    &       2.377    &       3.084      &     4.000    &       11.344 \\        
  K5    & 0.942      &     1.073     &      1.222    &       1.584     &      2.052    &       2.662    &       3.453      &     4.478    &       12.704 \\        
  K6    & 1.079      &     1.229     &      1.399    &       1.815     &      2.352    &       3.050    &       3.957      &     5.133    &       14.561 \\        
  K7    & 1.252      &     1.426     &      1.624    &       2.107     &      2.731    &       3.541    &       4.596      &     5.963    &       16.915 \\        
  M0    & 1.462      &     1.665     &      1.897    &       2.461     &      3.190    &       4.137    &       5.371      &     6.969    &       19.766 \\        
  M0.5  & 1.581      &     1.801     &      2.051    &       2.661     &      3.449    &       4.473    &       5.809      &     7.538    &       21.379 \\        
  M1    & 1.708      &     1.946     &      2.217    &       2.876     &      3.729    &       4.836    &       6.280      &     8.151    &       23.115 \\        
  M1.5  & 1.845      &     2.102     &      2.394    &       3.107     &      4.029    &       5.225    &       6.786      &     8.808    &       24.976 \\        
  M2    & 1.991      &     2.268     &      2.584    &       3.353     &      4.348    &       5.639    &       7.325      &     9.508    &       26.961 \\        
  M2.5  & 2.194      &     2.497     &      2.842    &       3.685     &      4.781    &       6.202    &       8.052       &    10.470    &      29.737 \\        
  M3    & 2.466      &     2.808     &      3.197    &       4.148     &      5.379    &       6.977    &       9.048       &    11.741    &      33.227 \\        
  M3.5  & 2.976      &     3.390     &      3.861    &       5.012     &      6.497    &       8.426    &       10.921      &    14.153    &      39.967 \\        
  M4    & 3.726      &     4.244     &      4.835    &       6.276     &      8.135    &       10.550    &      13.670      &    17.707    &      49.958 \\        
  M4.5  & 4.714      &     5.370     &      6.118    &       7.941     &      10.293    &      13.347    &      17.296      &    22.401    &      63.199 \\        
  M5    & 5.942      &     6.768     &      7.711    &       10.006     &     12.970    &      16.819    &      21.798      &    28.237    &      79.690 \\        
  M5.5  & 7.408      &     8.438     &      9.613    &       12.472     &     16.167    &      20.965    &      27.176      &    35.215    &      99.431 \\        
  M6    & 9.113      &     10.380     &     11.824    &      15.339     &     19.884    &      25.785    &      33.431      &    43.333    &      122.423 \\       
  M6.5  & 11.058      &    12.594     &     14.345    &      18.606     &     24.121    &      31.279    &      40.562      &    52.593    &      148.665 \\       
  M7    & 13.241      &    15.080     &     17.175    &      22.274     &     28.877    &      37.448    &      48.569      &    62.995    &      178.157 \\       
  M7.5  & 15.663      &    17.838     &     20.315    &      26.343     &     34.154    &      44.290    &      57.454      &    74.537    &      210.899 \\       
  M8    & 18.324      &    20.868     &     23.764    &      30.812     &     39.950    &      51.807    &      67.214      &    87.221    &      246.892 \\       
  M8.5  & 21.224      &    24.170     &     27.523    &      35.682     &     46.266    &      59.998    &      77.851       &   101.046    &     286.135 \\       
  M9    & 24.364      &    27.744     &     31.591    &      40.952     &     53.102    &      68.862    &      89.364       &   116.013    &     328.628 \\       
  M9.5  & 27.742      &    31.590     &     35.968    &      46.624     &     60.458    &      78.401    &      101.754      &   132.120    &     374.372 \\       
  L0    & 31.359      &    35.708     &     40.655    &      52.695     &     68.333    &      88.614    &      115.020      &   149.369    &     423.365 \\       

  \hline 
\end{tabular} 
\end{center} 
\end{table*}  
%%%%%%%%%%%%%%%%%%%%%%%%%%%%%%%%%%%%%%%%%%%%%%%%%%%%%%%%%%%%%%%%%%%%%%%%%%%%

\section{Plots of the spectra}

%%%%%%%%%%%%%%%%%%%%%%%%%%%%%%%%%%%%%
\begin{figure*}[!t]
\centering
\includegraphics[width=\textwidth]{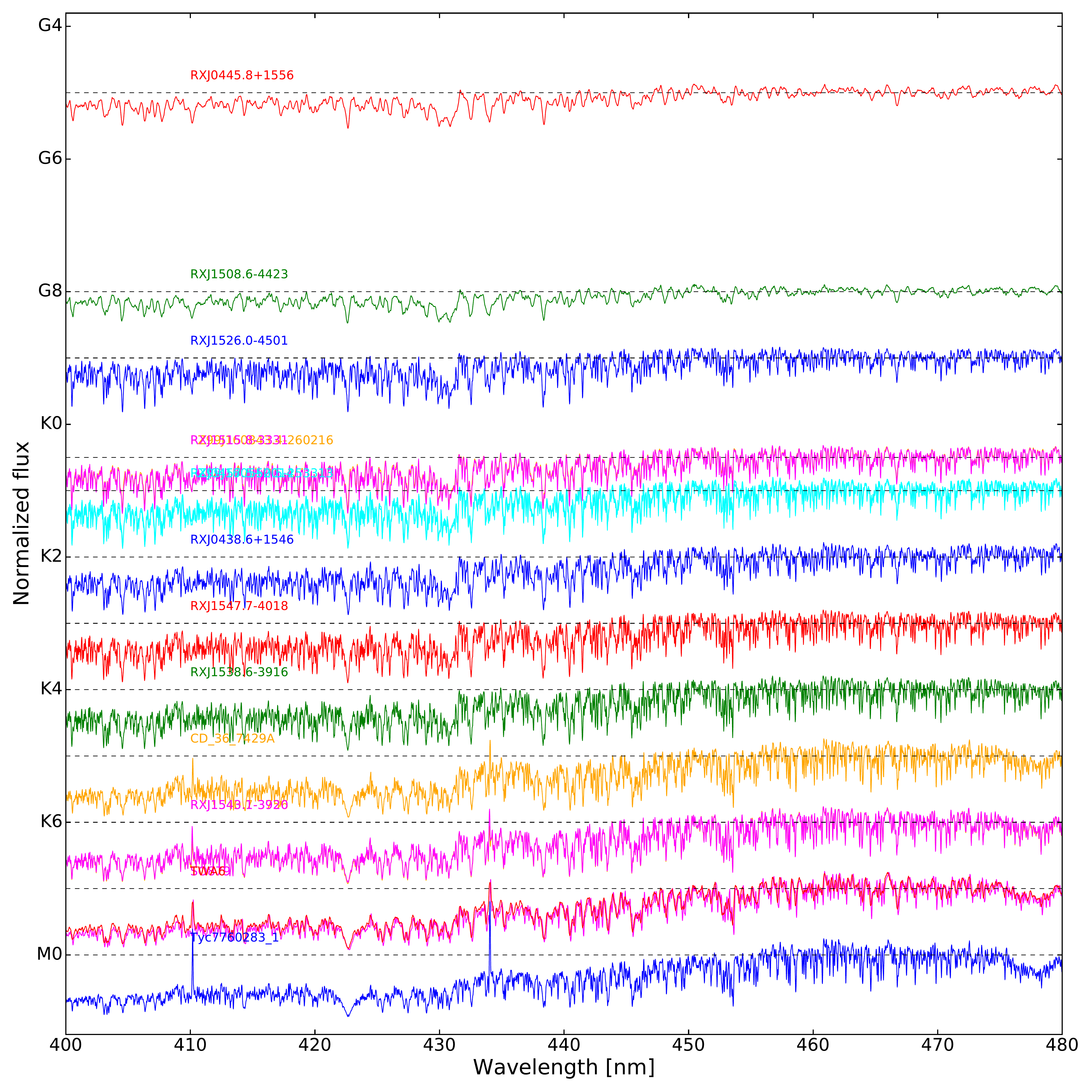}
\caption{Spectra of Class~III YSOs ordered according to their spectral type and normalized at 460 nm. 
     \label{fig::early1}}
\end{figure*}
%%%%%%%%%%%%%%%%%%%%%%%%%%%%%%%%%%%%%%%%%%%%%%%%%%%%%%%%%%%%%%%%%%%%%%%%%%%%

%%%%%%%%%%%%%%%%%%%%%%%%%%%%%%%%%%%%%
\begin{figure*}[!t]
\centering
\includegraphics[width=\textwidth]{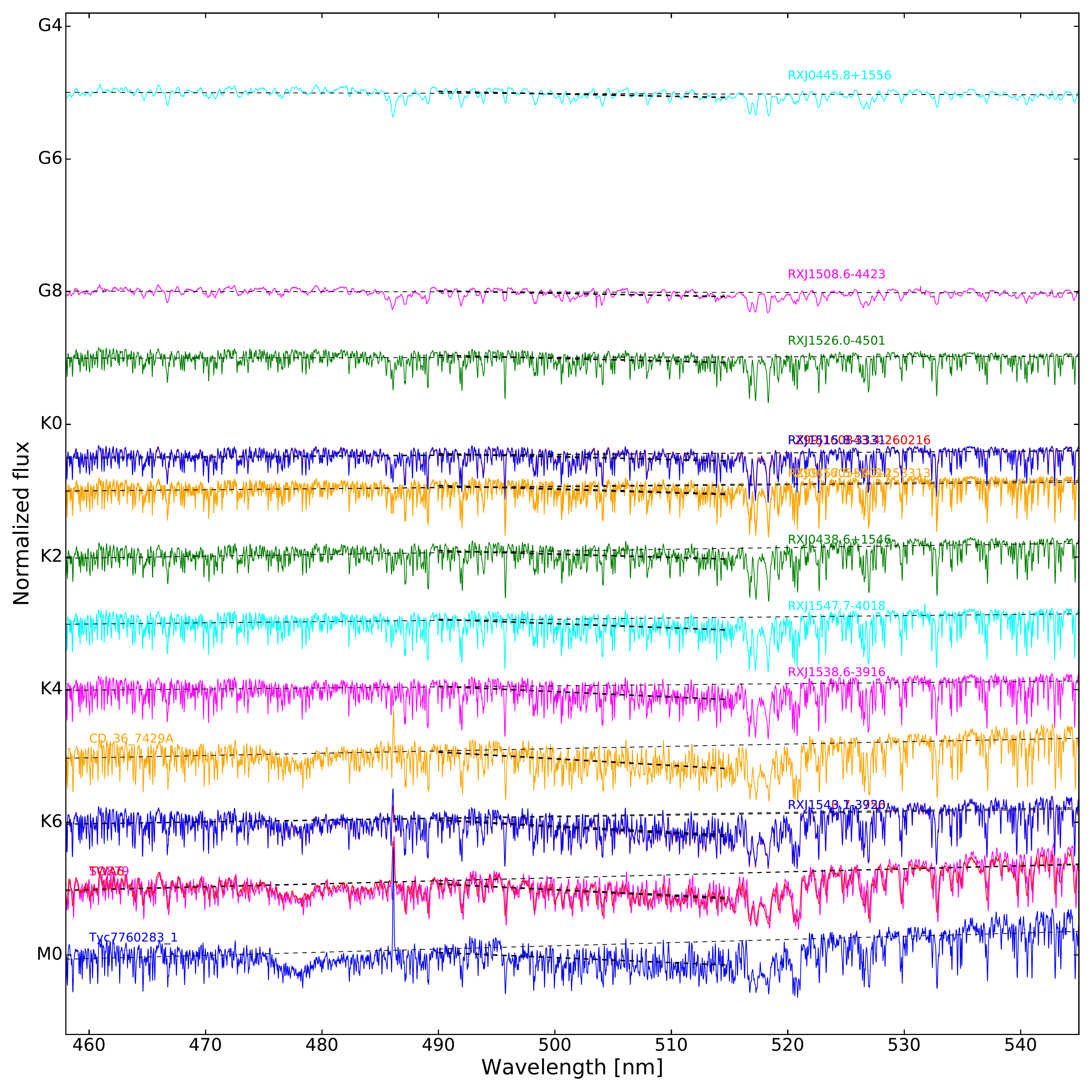}
\caption{Spectra of Class~III YSOs ordered according to their spectral type and normalized at 460 nm. 
     \label{fig::early2}}
\end{figure*}
%%%%%%%%%%%%%%%%%%%%%%%%%%%%%%%%%%%%%%%%%%%%%%%%%%%%%%%%%%%%%%%%%%%%%%%%%%%%

%%%%%%%%%%%%%%%%%%%%%%%%%%%%%%%%%%%%%
\begin{figure*}[!t]
\centering
\includegraphics[width=\textwidth]{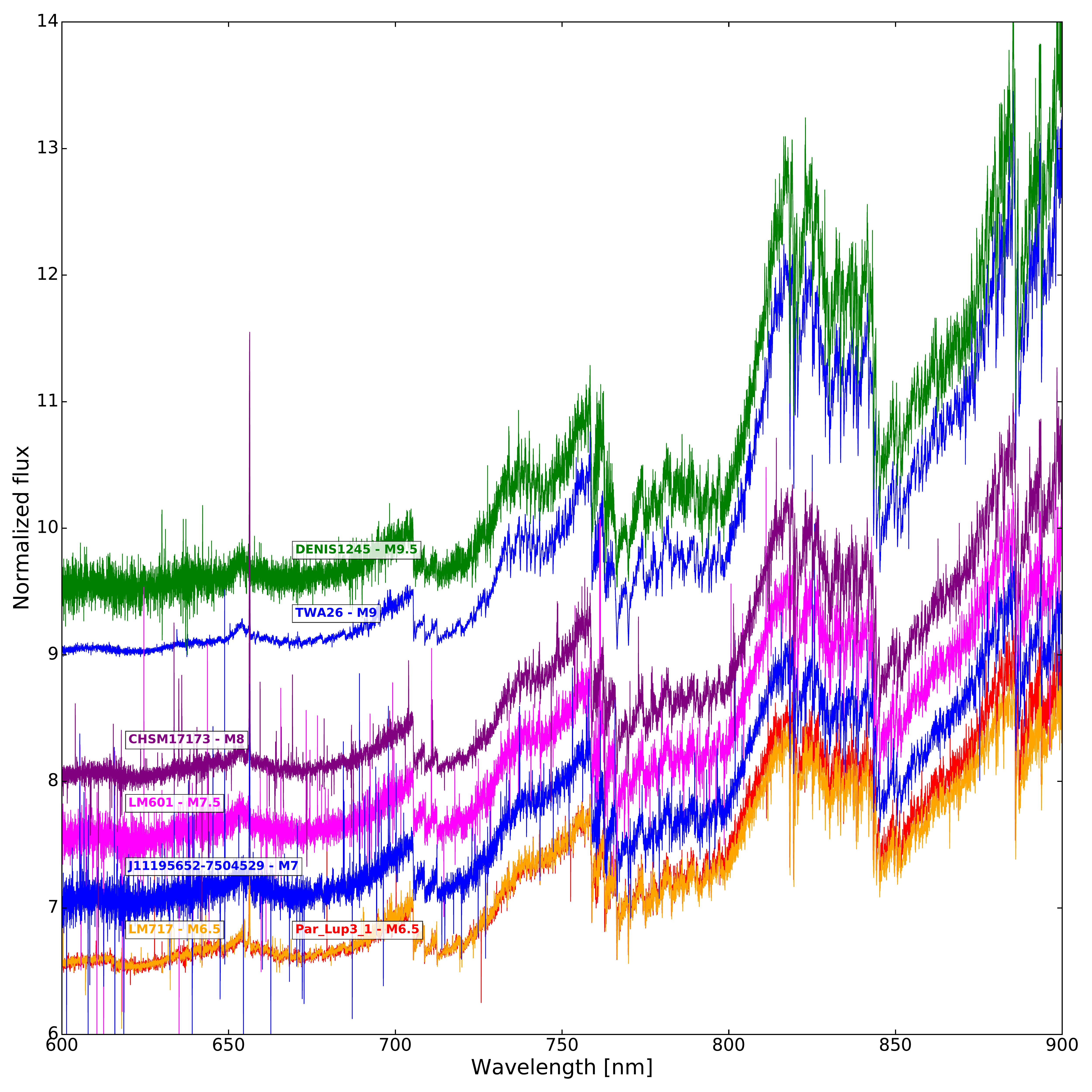}
\caption{Spectra of Class~III YSOs with spectral type later than M6 ordered according to their spectral type and normalized at 750 nm. 
     \label{fig::late}}
\end{figure*}
%%%%%%%%%%%%%%%%%%%%%%%%%%%%%%%%%%%%%%%%%%%%%%%%%%%%%%%%%%%%%%%%%%%%%%%%%%%%

\end{document}